\documentclass[AMA,STIX1COL]{WileyNJD-v2}
\articletype{Research article}%

\received{26 April 2016}
\revised{6 June 2016}
\accepted{6 June 2016}

\usepackage{float}
\usepackage{pmat}    
\usepackage{color}
\usepackage{epstopdf} 
\usepackage{graphicx}
\usepackage{amsfonts,amsmath}
\usepackage{multirow}
\usepackage{mathtools}
\usepackage[most]{tcolorbox}
\usepackage{xcolor}
\usepackage{adjustbox}
\usepackage{algorithm}
\usepackage{epsfig}
\usepackage{tabularx}
\usepackage{caption}
\usepackage{subcaption}
\usepackage{enumitem}
\usepackage{siunitx}
\usepackage{booktabs} % for \specialrule command
\usepackage{arydshln}

{}

\usepackage{calrsfs}
\DeclareMathAlphabet{\pazocal}{OMS}{zplm}{m}{n}
\newcommand{\Real}{\mathbb R}

\newcommand{\norm}[1]{\left\Vert#1\right\Vert}
\newcommand{\nx}{n_\mathrm{x}}
\newcommand{\dnu}{n_\mathrm{u}}
\newcommand{\np}{n_\mathrm{p}}
\newcommand{\ny}{n_\mathrm{y}}

\newcommand{\nnoise}{n}

\DeclareMathOperator*{\argmin}{arg\,min}

\newcommand{\idx}[2]{\mathbb{I}_{#1}^{#2}}

\newcolumntype{Y}{>{\centering\arraybackslash}X}

\begin{document}

\title{Bayesian Optimization Based Grid Point Allocation for LPV and Robust Control} %ler synthesis}

\author[1]{E. Javier Olucha*}
\author[2]{Arash Sadeghzadeh}
\author[1]{Amritam Das}
\author[1,4]{Roland  T{\'o}th}

\authormark{OLUCHA \textsc{et al.}}
\address[1]{\orgdiv{Control Systems Group},  \orgname{Eindhoven University of Technology}, \orgaddress{Eindhoven,  \country{The Netherlands}}}

\address[2]{\orgname{Data Science \& Artificial Intelligence},  \orgname{Breda University of Applied Sciences}, \orgaddress{Breda,  \country{The Netherlands}}}

\address[4]{\orgdiv{Systems and Control Laboratory}, \orgname{HUN-REN SZTAKI},  \orgaddress{Budapest,  \country{Hungary}}}

\corres{*E. Javier Olucha, \orgname{Eindhoven University of Technology}, \orgaddress{Eindhoven,  \country{The Netherlands}}, \email{e.j.olucha.delgado@tue.nl}.\vspace{-4mm}}

\abstract[Summary]{This paper investigates systematic selection of optimal grid points for grid-based \emph{linear parameter-varying} (LPV) and robust controller synthesis.
In both settings, the objective is to identify a set of local models such that the controller synthesized for these local models will satisfy global stability and performance requirements for the entire system.
Here, local models correspond to evaluations of the LPV or uncertain plant at fixed values of the scheduling signal or realizations of the uncertainty set, respectively.
Then, \emph{Bayesian optimization} is employed to discover the most informative points that govern the closed-loop performance of the designed LPV or robust controller for the complete system until no significant further performance increase or a user specified limit is reached.
Furthermore, when local model evaluations are computationally demanding or difficult to obtain, the proposed method is capable to minimize the number of evaluations and adjust the overall computational cost to the available budget.
Lastly, the capabilities of the proposed method in automatically obtaining a sufficiently informative grid set are demonstrated on three case-studies: a robust controller design for an unbalanced disk, a multi-objective robust attitude controller design for a satellite with uncertain parameters and two flexible rotating solar arrays, and an LPV controller design for a robotic arm.
}

\keywords{Linear parameter-varying systems, nonlinear systems, robust control, grid-based controller design, Bayesian optimization, machine learning.}

\jnlcitation{\cname{%
        \author{Olucha EJ}, \author{Sadeghzadeh A}, \author{Das A}, \author{T{\'o}th R}} (\cyear{2025}),
    \ctitle{Bayesian Optimization Based Grid Point Allocation for LPV and
Robust Control}, \cvol{2025}.}

\fundingInfo{European Space Agency (grant number: 4000145530) and The MathWorks Inc.\ Opinions, findings, conclusions or recommendations expressed in this abstract are those of the authors and do not necessarily reflect the views of The MathWorks Inc.\ or the European Space Agency.}
\maketitle

%%%%%%%%%%%%%%% MAIN MATTER %%%%%%%%%%%%%%%%%%%%%%%%%%%%%
\section{Introduction}\label{sec:intro}
The \emph{linear parameter-varying} (LPV) framework\cite{toth2010} is a well-established surrogate modeling approach for nonlinear systems. In LPV models, the signal relations between inputs and outputs are considered linear, but the underlying nonlinear dynamics are captured thanks to the introduction of the so-called \emph{scheduling variables}, denoted by $p$. While the LPV framework has been demonstrated to be capable of achieving high-performance controller design in numerous applications \cite{https://doi.org/10.1002/rnc.704,mohammadpourControlLinearParameter2012a,bachnasReviewDatadrivenLinear2014,hoffmannSurveyLinearParameterVarying2015}, one of the main challenges is obtaining suitable LPV models for the design process. Although numerous local or global LPV modeling techniques have been proposed\cite{kwiatkowskiAutomatedGenerationAssessment2006,decaignyInterpolationBasedModelingMIMO2011a,sadeghzadehAffineLinearParametervarying2020,koelewijnLearningReducedOrderLinear2023a,oluchaAutomatedLinearParameterVarying2025}, the resulting models are often too complex for the existing LPV controller synthesis tools. Typically, the produced models either depend affinely on scheduling variables at the expense of a high-dimensional scheduling space, or exhibit low-dimensional, but complicated scheduling dependencies. The former can be handled by scheduling dimension reduction techniques\cite{oluchaReductionLinearParameterVarying2024b}, whereas for the latter one must resort to grid-based LPV controller synthesis methods. These grid-based synthesis methods generally design either local controllers for a set of local model realizations followed by controller interpolation across the scheduling domain, or they discretize the global LPV synthesis problem, i.e., the semi-definite constraints of the synthesis problem, through gridding of the scheduling domain\cite{9568982, HJARTARSON2015139}.

%%%%%%%%%%%%%%%%%%%%%%%  ALTERNATIVE GRIDDED LPV JUSTIFICATION %%%%%%%%%%%%%%%%%%%%%%%
% When the LPV plant depends affinely on the scheduling variables, the synthesis problem can be addressed using polytopic methods\cite{APKARIAN19951251}. For rational dependencies, there are synthesis techniques based on full-block multipliers~\cite{SCHERER2001361} or D/G-scalings~\cite{scorletti98}. However, the numerical complexity of both polytopic and full-block multiplier approaches scales exponentially with the number of scheduling variables, while D/G-scaling methods trade reduced synthesis complexity at the expense of increased conservatism~\cite{hoffmannSurveyLinearParameterVarying2015}. When the number of scheduling variables is not small, the dependence on them is not rational, dynamic variations of model elements are represented via lookup tables, or the resulting controllers are overly conservative, grid-based LPV synthesis methods are used.
%%%%%%%%%%%%%%%%%%%%%%%%%%%%%%%%%%%%%%%%%%%%%%%%%%%%%%%%%%%%%%%%%%%%%%%%%%%%%%%%%%%%%

% Grid-based synthesis is also a common practice for practical applications of robust controllers. 
In the robust control synthesis framework\cite{zhouRobustOptimalControl1996, scherer2022robust}, \emph{linear fractional representations} (LFRs) are central in representing a wide range of uncertainties via the interconnection of a nominal linear system $P$ and an uncertainty block $\Delta$, together with control-loop configurations and performance shaping filters\cite{4789992}. Despite their demonstrated efficacy across multiple industrial domains, \cite{POSTLETHWAITE200727, liu2016robust}, scalability remains a critical challenge when dealing with large-scale systems. In such cases, the uncertainty description can involve numerous dynamic or parametric uncertainties, often with multiple repetitions, for which the existing software tools break down due to numerical scalability limitations and hardware constraints\cite{5612823}. To mitigate these issues, robust controller synthesis can be approached in a grid-based fashion by sampling of the involved uncertainty set, and a posteriori worst-case $\mu$-analysis\cite{packardComplexStructuredSingular1993} can be employed to certify robust stability and performance on the entire uncertainty set.

%%%%%%%%%%%%%%%%%%%%%%%  ALTERNATIVE GRIDDED ROBUST CONTROL JUSTIFICATION %%%%%%%%%%%%%%%%%%%%%%%
% Several well-established methods from robust control theory are available to address this problem, including the D/K-iteration\cite{balas94} and structured controller synthesis\cite{apkarianParametricRobustStructured2015} methods. However, their applicability is limited by uncertainty blocks $\mathbf{\Delta}$ with many uncertainty parameters or uncertainty parameter repetitions, where the synthesis becomes computationally intractable, or yields controllers with impractically high order.
% % Moreover, these synthesis methods do not provide guarantees on robust stability and performance of the closed-loop system. Then, and a posteriori $\mu$-analysis verification step is required to verify robustness, but the existing $\mu$-analysis tools break down due to numerical scalability issues. 
% In such cases, a grid-based robust controller synthesis approach can be employed.
%%%%%%%%%%%%%%%%%%%%%%%%%%%%%%%%%%%%%%%%%%%%%%%%%%%%%%%%%%%%%%%%%%%%%%%%%%%%%%%%%%%%%

As discussed above, grid-based approaches offer a practical alternative for LPV and robust controller design when system representations are too complex for the existing tools.
In both the LPV or robust control case, a crucial challenge is the strategic allocation of grid points, which should include the most informative points governing closed-loop stability and performance. Not many approaches address the grid point allocation problem. For example, gap-metric-based methods~\cite{theodoulisGainScheduledAutopilot2008,fleischmannSystematicLPVLFR2016a} can be employed to identify grid points where robustness margins are violated or to provide an indication for the gridding procedure in terms of grid density and structure. However, these approaches either rely on extensive line search, do not provide the precise point locations, or do not incorporate explicit closed-loop performance objectives. 
Another example, applicable only in the robust control case, relies on $\mu$-analysis to identify worst-case uncertainty realizations that degrade closed-loop performance and to iteratively refine the grid accordingly. Although the iterative application of $\mu$-analysis and controller synthesis can successfully identify key uncertainty points, the computational burden is substantial. Moreover, $\mu$-analysis struggles with moderate- to high-dimensional uncertainty blocks, which is particularly limiting in applications such as aerospace systems. For example, a satellite with two rotating appendages already poses a challenge for current $\mu$-analysis~\cite{sanfedinoExperimentalValidationHigh2019a}.

As a result, grid-point allocation is often guided by engineering insight and human-in-the-loop iterative procedures. This process can become tedious and time-consuming, delaying the development cycle of new prototypes, particularly when evaluating or sampling LPV or uncertain LFR models entails high computational costs.
Not all the grid-points are equally informative with respect to the closed-loop stability and performance requirements, and the naive addition of grid points does not necessarily yield a better representation of the entire dynamics.
Moreover, in the absence of systematic guidelines on adjusting the grid resolution, the resulting grids are often unnecessarily dense in some regions of the scheduling/uncertainty space, increasing the computational load of the synthesis, or too sparse in other regions where critical performance changes happen, potentially degrading the closed-loop performance or even compromising the stability of the real system.
Consequently, a key problem in grid-based LPV and robust controller synthesis is the lack of a general, systematic, and computationally efficient grid-point allocation procedure that explicitly considers closed-loop performance requirements, scales to high-dimensional scheduling or uncertainty spaces, and minimizes the number of required system evaluations.

In this paper, we propose a systematic grid-point allocation method for grid-based LPV and robust controller synthesis. The core idea is to define a cost function that quantifies the informativity of a candidate grid-point with respect to the desired closed-loop performance objectives. To efficiently optimize this function, we employ \emph{Bayesian optimization} (BO)~\cite{shahriariTakingHumanOut2016a,Frazier:2018iqo}, which is particularly suited for the nature of our problem due to its efficiency in solving high-dimensional, nonlinear optimization problems,
% while minimizing the number of function evaluations, 
even when the explicit form of the cost function is not available~\cite{brochu2010,brosigDataefficientAutotuningBayesian2020}.
We use BO to maximize the  informativity cost function, i.e., to identify the grid point that yields the most informative local model for the closed-loop performance objectives. Once the most informative grid point is found, the corresponding local model is included in the controller design. This process is iterated either until a predefined number of grid points is reached or no further significant improvement in the closed-loop performance objectives is observed. Additionally, the method can be adjusted based on the computational budget available as both the number of cost function evaluations and local model queries can be limited at each iteration.

This paper is structured as follows. First, we introduce the grid-point allocation problem for LPV and robust controller synthesis in Section~\ref{Sec:Problem_Statement}. Then, the Bayesian optimization-based grid-point allocation strategy is proposed in Section~\ref{Sec:method}. In Section~\ref{Sec:simulation}, we demonstrate and analyse the performance of the proposed method on robust controller design for an unbalanced disk system, a multi-objective robust controller design for a satellite with uncertain parameters and two flexible rotating appendages, and an LPV controller design for a two-degree-of-freedom robotic arm. Finally, the conclusions about the proposed approach are drawn in Section~\ref{Sec:conclusion}.

\noindent
\textbf{Notation:} The set of real numbers is denoted by $\Real$. The set of positive integers is denoted by $\mathbb{Z^+}$. For sets $\mathbb{A}$ and $\mathbb{B}$, $\mathbb{B}^{\mathbb{A}}$ indicates the collection of all maps from $\mathbb A$ to $\mathbb B$. The set of all proper, stable and real rational transfer matrices is denoted as $\pazocal{RH}_\infty$. The vector $\left(\ x_1^\top \ \cdots \ x_n^\top \ \right)^\top$ is composed by $\operatorname{col}\left(x_1, \, \dots, \, x_n \right)$. 
% Given the real, rational and proper transfer matrices $\Delta : \mathbb{C} \to \mathbb{C}^{n_\mathrm{d} \times m_\mathrm{d}}$, $P: \mathbb{C} \to \mathbb{C}^{n \times m}$ and $K : \mathbb{C} \to \mathbb{C}^{n_\mathrm{k} \times m_\mathrm{k}}$ with $n_\mathrm{k}, n_\mathrm{d} < m$ and $m_\mathrm{k}, m_\mathrm{d} < n$, let $\pazocal{F}_\mathrm{u} \left(P, \Delta \right)$, $\pazocal{F}_\mathrm{l} \left(P, K \right)$ and $\pazocal{F}_\mathrm{bi} \left(\Delta, P, K \right)$ denote the upper, the lower and the bidirectional linear fractional transformation, respectively.
\section{Problem statement} \label{Sec:Problem_Statement}
\begin{figure}[b]
    \centering
    % \vspace{-4pt}
    \includegraphics[width=0.40\columnwidth]{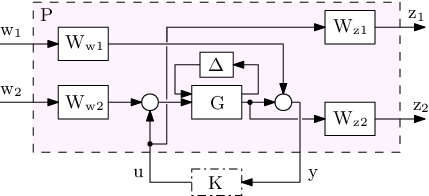}
    % \vspace{-5pt}
    \caption{Example of a generalized plant $P$, containing an uncertain system $G$, the generalized disturbance channels $w_{\{\bullet\}}$, the generalized performance channels $z_{\{\bullet\}}$, the weighting filters $W_{\{ \bullet\}}$ that capture the desired closed-loop behaviour specifications and the to-be-synthesized controller $K$.}
    \label{fig:genplant}
    % \vspace{-17pt}
\end{figure}

% The generalized plant concept is a systematic approach for analysis and controller synthesis for uncertain LTI and LPV models. Through this framework, a wide range of plant and control configurations can be described, where performance specifications can be represented and closed-loop behaviour can be shaped with weighting filters in an intuitive manner. 
In this work, we consider the generalized plant concept to formulate control configurations and desired performance specifications. An example of a generalized plant connected with a controller $K$ is depicted in Fig.~\ref{fig:genplant}.
To address a rather general class of controller design problems consider the following system representation given by
\begin{equation}
  \begin{pmatrix}
                  z_\Delta \\ z \\ y
              \end{pmatrix} =  \underbrace{\begin{pmatrix}
                  G_{\mathrm{d d}} & G_{\mathrm{d w}} & G_{\mathrm{d u}} \\
                  G_{\mathrm{z d}} & G_{\mathrm{z w}} & G_{\mathrm{z u}} \\
                  G_{\mathrm{y d}} & G_{\mathrm{y w}} & G_{\mathrm{y u}}
              \end{pmatrix}}_{G}
              \begin{pmatrix}
                  w_\Delta \\ w \\ u
              \end{pmatrix}, \quad w_\Delta = \Delta z_\Delta,
\end{equation}
where $u: \mathbb{R} \rightarrow \mathcal{U} \subseteq \mathbb{R}^{n_\mathrm{u}}$ is the control input, $y: \mathbb{R} \rightarrow \mathcal{Y} \subseteq \mathbb{R}^{n_\mathrm{y}}$ is the measured output, $w:\mathbb{R} \rightarrow \pazocal{W} \subseteq \mathbb{R}^{n_\mathrm{w}}$ corresponds to the generalized disturbance channels (consisting of references, disturbances, etc.), \mbox{$z:\mathbb{R} \rightarrow \pazocal{Z} \subseteq \mathbb{R}^{n_\mathrm{z}}$} to the generalized performance channels (consisting of tracking errors, control efforts, etc.), $w_\Delta:\mathbb{R} \rightarrow \pazocal{W}_\Delta \subseteq \mathbb{R}^{n_\mathrm{q}}$ to the input uncertainty channels, \mbox{$z_\Delta:\mathbb{R} \rightarrow \pazocal{Z}_\Delta \subseteq \mathbb{R}^{n_s}$} to output uncertainty channels,
and $G:(\mathbb{R}^{n_\mathrm{d}+n_\mathrm{w}+n_\mathrm{u}})^\mathbb{R} \rightarrow (\mathbb{R}^{n_\mathrm{d}+n_\mathrm{z}+n_\mathrm{y}})^\mathbb{R}$ and $\Delta:(\mathbb{R}^{n_\mathrm{d}})^\mathbb{R} \rightarrow (\mathbb{R}^{n_\mathrm{d}})^\mathbb{R}$ are causal operators with $G$ being \emph{linear and time invariant} (LTI). 

$G$ can also be represented in terms of a continuous-time state-space representation in the form of :
\begin{equation}
    \begin{pmatrix}
                 \dot{x}(t) \\ z_\Delta(t)  \\ z(t) \\ y(t)
             \end{pmatrix} = \underbrace{\begin{pmatrix}
                            A &        B_\mathrm{d}                    & B_\mathrm{w}     & B_\mathrm{u}     \\
                              C_\mathrm{d} & D_\mathrm{dd} & D_{\mathrm{d w}} & D_{\mathrm{d u}} \\
                   C_\mathrm{z} & D_\mathrm{zd} & D_\mathrm{zw}    & D_\mathrm{zu}    \\
                   C_\mathrm{y} & D_\mathrm{yd}  & D_\mathrm{yw}    & D_\mathrm{yu}    \\
              \end{pmatrix}}_M \begin{pmatrix}
                 x(t) \\ w_\Delta(t) \\ w(t) \\ u(t)
              \end{pmatrix},
\end{equation}
where $x : \mathbb{R} \to \mathcal{X} \subseteq \mathbb{R}^{\nx}$ is the state variable and $(A, B_\mathrm{d}, \ldots D_\mathrm{yu})$ are matrices over $\mathbb{R}$ with appropriate dimensions. Equivalently, due to its LTI nature, $G$ can be represented in a transfer function form:
\begin{equation}
    G(s)= \begin{pmatrix} D_{\mathrm{d d}} & D_{\mathrm{d w}} & D_{\mathrm{d u}} \\ D_{\mathrm{z d}} & D_{\mathrm{z w}} & D_{\mathrm{z u}}
    \\ D_\mathrm{yd} & D_\mathrm{yw}    & D_\mathrm{yu}   \end{pmatrix} + \begin{pmatrix} C_{\mathrm{d}} \\ C_{\mathrm{z}} \\ C_\mathrm{y}\end{pmatrix} 
    \left(I s - A \right)^{-1} 
    \begin{pmatrix}
        B_{\mathrm d} & B_{\mathrm w} & B_{\mathrm u}
    \end{pmatrix} %, \quad s \in \mathbb{C},
\end{equation}
with $s\in\mathbb{C}$ the complex frequency. Here, with a slight abuse of notation, we denote by $G(s)$ the transfer function, i.e.,  $G\in \mathbb{R}^{(n_\mathrm{d}+n_\mathrm{z}+n_\mathrm{y})\times (n_\mathrm{d}+n_\mathrm{z}+n_\mathrm{y})}[s]$, where $\mathbb{R}^{n \times m}[s]$ is the set of real-rational and proper $(n \times m)$-matrix-valued functions in $s$.
Depending on the nature of the delta block $\Delta$, we consider and distinguish two different cases:
\begin{itemize}
    \item Uncertain LTI models:
    \begin{equation}
        w_\Delta (t) = \Delta z_\Delta(t),
    \end{equation}
    where $\Delta$ represents dynamic variations due to uncertain parameters of the system, defined as
    \begin{equation} \label{deltaSet}
              \Delta \in \mathbf{\Delta} \coloneq  \left\{ \Delta \in \Real^{n_{\mathrm{s}} \times n_{\mathrm{q}}} \ | \ \Delta \in \mathbf{V}_\mathrm{c}, \|\Delta\|_{2,2} < 1
              \right\},
          \end{equation}
          with the  structured value set $\mathbf{V}_\mathrm{c}$ given by
          \begin{gather} \label{VSet}
              \mathbf{V}_\mathrm{c} :=  \left\{ V = \operatorname{diag} \left(
              \eta_1 I_{n_1}, \dots, \eta_{n_{\mathrm r}}I_{n_{\mathrm r}}
              \right) \ | \ \eta_i \in \Real \right\},
          \end{gather}
          such that $\eta_i$ represents a parametric uncertainty with $\mathrm{n_r}\geq 1$ repetitions and $\|\bullet\|_{2,2}$ corresponds to the $2,2$ matrix norm.
          % , i.e., the maximum singular value.
    \item LPV models:
    \begin{equation}
        w_\Delta (t) = \Delta (p (t)) z_\Delta(t),
    \end{equation}
    where 
    $p(t) \in \mathbb{P}$
    % $p : \mathbb{R} \to \mathcal{P} \subseteq \mathbb{R}^{n_{\mathrm p}}$ 
    is the collection of scheduling variables that represent external effects that induce dynamic variations or nonlinear dynamic relationships, and $\mathbb{P} := \{\underline{p}_i \le p_i (t) \le \overline{p}_i\}_{i=1}^{n_\mathrm{p}}$
    % and the compact set $\mathbb{P \subseteq \mathcal{P}}$, $\mathbb{P} := \{\underline{p}_i \le p_i (t) \le \overline{p}_i\}_{i=1}^{n_\mathrm{p}}$ 
    is the admissible scheduling signal space where $\underline{p}_i$ and $\overline{p}_i$ are the respective lower and upper bounds. $\Delta$ is assumed to have the following diagonal structure:
     \begin{equation} \label{eq:deltaLPVLFR}
              \Delta(p(t)) = \operatorname{diag} \left(
              p_1(t) I_{n_1}, \dots, p_{n_ \mathrm p}(t)I_{n_ \mathrm p} \right).
          \end{equation}
\end{itemize}
In the uncertain LTI case, by applying the Redheffer star product, denoted by $\star$, on the transfer function representation (corresponding to the upper fractional transform), we get the uncertain transfer function representation:
\setlength{\dashlinegap}{1pt}
\begin{equation} \label{eq:genplantRobust}
    P_{\Delta}:\biggl\{\ \  \Delta \star G(s) = \begin{pmatrix}
        G_{\mathrm{zw}}(s) & G_{\mathrm{zu}}(s) \\ G_{\mathrm{yw}}(s) & G_{\mathrm{yu}}(s)
    \end{pmatrix} + 
    \begin{pmatrix}
        G_{\mathrm{zd}}(s) \\ G_{\mathrm{yd}}(s)
    \end{pmatrix} \Delta \left(I - G_{\mathrm{dd}}(s) \Delta \right)^{-1} 
    \begin{pmatrix}
        G_{\mathrm{dw}}(s) & G_{\mathrm{du}}(s)
    \end{pmatrix} := \left(\begin{array}{c:c}
       P_{\mathrm{zw}}  & P_{\mathrm{zu}} \\
       \hdashline
       P_{\mathrm{yw}}  & P_{\mathrm{yu}}
    \end{array}\right)(s,\Delta).
\end{equation}
Similarly we can compute the uncertain plant representation in terms of a state-space form.
In the LPV case, in absence of a transfer function representation of the resulting operator,  by applying the star product on the SS representation, we obtain:
\begin{subequations}\label{eq:genplantLPVLFR} \begin{equation}
       P_{p}:\biggl\{\ \  \Delta(p) \star \Breve{M} = \begin{pmatrix}
        A & B_{\mathrm{w}} & B_{\mathrm{u}} \\
        C_{\mathrm{z}} & D_{\mathrm{zw}} & D_{\mathrm{zu}} \\
        C_{\mathrm{y}} & D_{\mathrm{yw}} & D_{\mathrm{yu}}
    \end{pmatrix} + 
    \begin{pmatrix}
        B_{\mathrm{d}} \\ D_{\mathrm{zd}} \\ B_{\mathrm{yd}}
    \end{pmatrix} \Delta(p) \left(I - D_{\mathrm{dd}} \Delta(p)\right)^{-1}
    \begin{pmatrix}
        C_{\mathrm{d}} & D_{\mathrm{dw}} & D_{\mathrm{du}}
    \end{pmatrix}=\bar{M}(p),
\end{equation}
where
\begin{equation}
 \begin{pmatrix}
                 \dot{x}(t) \\ z(t) \\ y(t)
             \end{pmatrix} = \underbrace{\begin{pmatrix}
                            \bar{A}(p) &         \bar{B}_\mathrm{w}(p)     & \bar{B}_\mathrm{u}(p)     \\
                             \bar{C}_\mathrm{z}(p) & \bar{D}_\mathrm{zw}(p)    & \bar{D}_\mathrm{zu}(p)    \\
                   \bar{C}_\mathrm{y}(p) & \bar{D}_\mathrm{yw}(p)    & \bar{D}_\mathrm{yu}(p)    \\
              \end{pmatrix}}_{\bar{M}(p)} \begin{pmatrix}
                 x(t) \\ w(t) \\ u(t)
              \end{pmatrix} \quad \text{and} \quad
    \breve{M} = \begin{pmatrix}
        0 & I_{n_\mathrm{d}} & 0 & 0 \\ I_{n_\mathrm{x}}  & 0 & 0 & 0 \\ 0 & 0 & I_{n_\mathrm{z}}  & 0 \\ 0 & 0 & 0 & I_{n_\mathrm{y}} 
    \end{pmatrix} \begin{pmatrix}
                            A &        B_{d}                    & B_\mathrm{w}     & B_\mathrm{u}     \\
                              C_\mathrm{d} & D_\mathrm{dd} & D_{\mathrm{d w}} & D_{\mathrm{d u}} \\
                   C_\mathrm{z} & D_\mathrm{zd} & D_\mathrm{zw}    & D_\mathrm{zu}    \\
                   C_\mathrm{y} & D_\mathrm{yd}  & D_\mathrm{yw}    & D_\mathrm{yu}    \\
              \end{pmatrix} \begin{pmatrix}
        0 & I_{n_\mathrm{d}}  & 0 & 0 \\ I_{n_\mathrm{x}}  & 0 & 0 & 0 \\ 0 & 0 & I_{n_\mathrm{w}}  & 0 \\ 0 & 0 & 0 & I_{n_\mathrm{u}} 
    \end{pmatrix},
\end{equation}
\end{subequations}
which is an LPV state-space representation with rational dependence of the matrix coefficients on $p\in\mathcal{P}=\mathbb{P}^\mathbb{R}$ as a signal. Note that if $D_{\mathrm{dd}}$ is zero, then~\eqref{eq:genplantLPVLFR} corresponds to LPV-SS representations with affine dependence on the scheduling variable:
\begin{equation}\label{eq:genplantLPV}
     P_{p}:\biggl\{\ \  \Delta(p) \star \Breve{M} = \begin{pmatrix}
        A & B_{\mathrm{w}} & B_{\mathrm{u}} \\
        C_{\mathrm{z}} & D_{\mathrm{zw}} & D_{\mathrm{zu}} \\
        C_{\mathrm{y}} & D_{\mathrm{yw}} & D_{\mathrm{yu}}
    \end{pmatrix} + 
    \begin{pmatrix}
        B_{\mathrm{d}} \\ D_{\mathrm{zd}} \\ B_{\mathrm{yd}}
    \end{pmatrix} \Delta(p)
    \begin{pmatrix}
        C_{\mathrm{d}} & D_{\mathrm{dw}} & D_{\mathrm{du}}
    \end{pmatrix}=\bar{M}(p).
\end{equation}
For the ease of notation, we can also write  
\begin{equation} \label{eq:LPV:op}
P_p = \Delta(p) \star G = \left[\begin{array}{c|cc} \bar{A}(p) &         \bar{B}_\mathrm{w}(p)     & \bar{B}_\mathrm{u}(p)     \\ \hline
                             \bar{C}_\mathrm{z}(p) & \bar{D}_\mathrm{zw}(p)    & \bar{D}_\mathrm{zu}(p)    \\
                   \bar{C}_\mathrm{y}(p) & \bar{D}_\mathrm{yw}(p)    & \bar{D}_\mathrm{yu}(p)    \end{array} \right]
\end{equation} 
corresponding to the state-space representation \eqref{eq:genplantLPVLFR} defined operator $P_p$.

Additionally, a rather general class of nonlinear systems can be represented %with general dependence of the resulting state-space matrices on $p(t)$. This case can be handled 
by allowing $\Delta$ to be any function of $p$ instead of the diagonal linear structure of \eqref{eq:deltaLPVLFR}. Furthermore, we assume that all weighting filters and signal connections are already contained in the plant description.

Now, for a given uncertain LTI model $P_\Delta$, the general controller synthesis problem is formulated as
\begin{subequations}\label{eq:robustsynthesis}
    \begin{align}
        K^\ast =&  \argmin_{K} \sup_{\Delta \in {\bf \Delta}} J(P_\Delta,  K)  \quad \text{subject to}                                         \\
                 &P_\Delta \star K \text{\ is internally stable }  
                \forall\Delta \in \mathbf{\Delta},
    \end{align}
\end{subequations}
where the closed-loop performance cost $J : (P_\Delta, K) \mapsto \gamma \in \Real$ is typically defined as the $\mathcal{H}_\infty$ or $\mathcal{H}_2$ norm of the closed-loop system, the controller $K$ is considered in the form
\begin{equation}\label{eq:robustController}
    K : \biggl\{ \ \begin{aligned}
        \dot{x}_{\mathrm{K}} &= A_{\mathrm{K}} x_{\mathrm{K}} + B_{\mathrm{K}} y, \\
        u &= C_{\mathrm{K}} x_{\mathrm{K}} + D_{\mathrm{K}} y,
    \end{aligned}
\end{equation}
and $P_\Delta \star K$ is internally stable if the following transfer matrix is stable:
\begin{equation}\label{eq:9blocktest}
    \begin{pmatrix}
        P_{\mathrm{zw}}(\Delta) & P_{\mathrm{zu}}(\Delta) & 0 \\
        0 & I & 0 \\
        P_{\mathrm{yw}}(\Delta) & P_{\mathrm{yu}}(\Delta) & I
    \end{pmatrix} + \begin{pmatrix}
        P_{\mathrm{zu}}(\Delta) \\ I \\ P_{\mathrm{yu}}(\Delta)
    \end{pmatrix} K \begin{pmatrix}
        I - P_{\mathrm{yu}}(\Delta)
    \end{pmatrix}^{-1} \begin{pmatrix}
        P_{\mathrm{yw}}(\Delta) & P_{\mathrm{yu}}(\Delta) & I
    \end{pmatrix}.
\end{equation}
For a given LPV model $P_{p}$, the general LPV controller synthesis problem is formulated as
\begin{subequations}\label{eq:LPVsynthesis}
    \begin{align}
     K_{p}^\ast =& \argmin_{K_{\mathrm{p}}}  \sup_{p \in {\mathcal{P}}} J(P_\mathrm{p},  K_{\mathrm{p}})  \quad \text{subject to}\\
    &P_{p} \star K_{p} \text{ is internally stable } \forall p \in \mathcal{P},
    \end{align}
\end{subequations}
where the closed-loop performance cost $J : (P_p, K_{\mathrm{p}}) \mapsto \gamma \in \Real$ is typically defined as the induced $\pazocal{L}_2$-gain or generalized $\mathcal{H}_2$ norm of the closed-loop system, the LPV controller $K_{p}$ is considered in the form
\begin{equation}\label{eq:LPVcontroller}
    K_{p} : \biggl\{ \ \begin{aligned}
        \dot{x}_{\mathrm{K}} &= A_{\mathrm{K}}(p) x_{\mathrm{K}} + B_{\mathrm{K}}(p) y, \\
        u &= C_{\mathrm{K}}(p) x_{\mathrm{K}} + D_{\mathrm{K}}(p) y,
    \end{aligned}
\end{equation}
where $A_{\mathrm K}: \mathbb{P}\to \Real^{n_{\mathrm{x_k}} \times n_{\mathrm{x_k}}}$, $B_{\mathrm K}: \mathbb{P}\to \Real^{n_{\mathrm{x_k}} \times n_{\mathrm{u_k}}}$, $C_{\mathrm K}: \mathbb{P}\to \Real^{n_{\mathrm{y_k}} \times n_{\mathrm{x_k}}}$ and $D_{\mathrm K}: \mathbb{P}\to \Real^{n_{\mathrm{y_k}} \times n_{\mathrm{u_k}}}$ are bounded matrix functions, and $P_{p} \star K_{p}$ is internally stable if there exists a matrix function $X:\mathbb{P} \to \Real^{(n_{\mathrm{x} }+n_{\mathrm{x_k}}) \times (n_{\mathrm{x} }+n_{\mathrm{x_k}})}$  with $X \in \mathcal{C}_1$ such that
\begin{equation}\label{eq:LPVinternalstability}
{A}_{\mathrm{cl}}^\top(\mathrm{p}) X(\mathrm{p}) + X(\mathrm{p}) {A}_{\mathrm{cl}}(\mathrm{p}) + \sum_{k=1}^{n_{\mathrm{p}}}\frac{\partial X(\mathrm{p})}{\partial \mathrm{p}_k} \mathrm{v}_k \prec 0, \quad X(\mathrm{p}) \succ 0, \quad \forall \mathrm{p}\in\mathbb{P}, \quad \forall\mathrm{v}\in\mathbb{V},
\end{equation}
where $\mathrm{p}$ corresponds to constant vectors, i.e., values that $p(t)\in\mathbb{P}$ can take and $\mathrm{v}\in\mathbb{V}$ corresponds to values that $\dot{p}(t)$ can take with a value set $\mathbb{V}\subseteq \mathbb{R}^{n_\mathrm{p}}$. Furthermore,
${A}_{\mathrm{cl}}(p)$ is defined by the closed-loop interconnection $P_{p} \star K_{p}$ in terms of
\begin{equation}
P_{p} \star K_{p}
=\left[\begin{array}{c|c}
A_{\mathrm{cl}}(p) & B_{\mathrm{cl}}(p) \\
\hline C_{\mathrm{cl}}(p) & D_{\mathrm{cl}}(p)
\end{array}\right].
\end{equation}
Note that the star-product-based construction of $A_{\mathrm{cl}}(p)$ results in a non-minimal state-space representation that has an essential role to imply internal stability of the interconnection, see \cite{Zhou95}.

%%%%%%%%
The main difference between~\eqref{eq:robustsynthesis} and~\eqref{eq:LPVsynthesis} lies in the form of  (i) the internal stability condition and (ii) performance and stability conditions that are needed to be satisfied for different type of variations of $\Delta$. In the robust synthesis problem, stability must be guaranteed for all $\Delta \in \mathbf{\Delta}$, i.e., in point-wise sense. In contrast, the LPV synthesis requires stability and performance to hold along all admissible scheduling \emph{trajectories} $p\in\mathcal{P}=\mathbb{P}^\mathbb{R}$. The latter problem is resolved by realising that both internal stability and popular performance objectives such as $\mathcal{L}_2$-gain can be expressed with stability and dissipativity constraints that are dependent on only the possible values that $p(t)$ and $\dot{p}(t)$ can take, such as in \eqref{eq:LPVinternalstability}, see\cite{briat2015}. This means that similar to the robust synthesis case, the problem can be re-cast to satisfying conditions for $P_\mathrm{p} \star K_\mathrm{p}$ with $p\in\mathbb{P}$ under all $v\in\mathbb{V}$ as constant vectors, see\cite{hoffmannSurveyLinearParameterVarying2015a}.

However, as detailed in Section~\ref{sec:intro}, solving~\eqref{eq:robustsynthesis} and~\eqref{eq:LPVsynthesis} is often computationally intractable for high state and scheduling dimensions or complicated dependency structures, where implementation of  exact LMI-based synthesis methods, e.g., polytopic approaches, full-block multiplier or IQC methods simply run out of memory or the solvers do not converge due to numerical reasons. Therefore, in practice, real-world design problems are often addressed in a grid-based fashion, where the uncertainty or scheduling space are discretized into a \emph{finite} set of points, and the synthesis procedures are reduced to satisfying the stability and performance constraints only at the grid locations. Hence, instead of a global synthesis problem, the design reduces to a multimodel synthesis problem with respect to LTI systems realizations at a discretized space of variations. 
 This allows to treat  the robust and LPV synthesis problems within a unified framework.

While the gridded formulation enables computational tractability, its effectiveness critically depends on the chosen grid points. The main challenge thus lies in determining the best selection of points such that, when the gridded synthesis is accomplished, the resulting controller will satisfy the global performance and stability specifications. In this work, our focus is not on the synthesis problems themselves, but rather on the grid point allocation problem:

For a given uncertain model $P_\Delta$ or LPV model $P_{p}$ and a given cost function $J$, determine the selection $\Theta := \{\theta^{(i)}\}_{i}^{N_\theta} \subset \Xi$ containing  $N_{\theta} \in \mathbb{Z}^+$ distinct points $\theta$ such that solving the grid-based controller synthesis problem defined as:
\begin{subequations}\label{eq:grid_synth}
\begin{align}
 K^\ast =& \argmin_K \max_{\theta \in \Theta} J(P_\theta, K_\theta) \quad \text{subject to} \\
        &P_\theta\star K_\theta \text{ is internally stable } \forall\theta \in \Theta,
    \end{align}
\end{subequations}
the corresponding controller will achieve $\sup_{\theta \in \Theta} J(P_\theta,K_\theta) \leq \gamma$ and internal stability of $P_\theta \star K_\theta$ for all $\Xi$, i.e., a given $\gamma\geq 0$ satisfaction of the global performance and stability objectives.

In case of a robust synthesis problem, $\theta := (\eta_1 \ \cdots \ \eta_{n_\mathrm{r}})^\top$, $\Delta(\theta) =  \operatorname{diag} \left(
              \eta_1 I_{n_1}, \dots, \eta_{n_{\mathrm r}}I_{n_{\mathrm r}}
              \right)$,  $\Xi := \mathbf{\Delta}$ and the controller K is of the form~\eqref{eq:robustController}. In case of an LPV synthesis problem, $\theta := p$, $\Xi := \mathbb{P}$, and the controller K is of the form~\eqref{eq:LPVcontroller} while $X$ is considered to be independent of $p$. Moreover, for a fixed static value $\theta \in \Xi$, any of the considered model representations \eqref{eq:genplantRobust} and \eqref{eq:genplantLPVLFR}
  correspond to an LTI system, denoted by $P_\theta := \Delta(\theta) \star G$ and defined in terms of \eqref{eq:genplantRobust} for a given uncertain model and \eqref{eq:LPV:op} for the LPV case with a constant $\mathrm{p}$. The resulting LTI models correspond to the local dynamics of the system. Then, the set $P_{\Theta} := \{P_\theta\}_{i=1}^{N_\theta}$ represents the collection of LTI models corresponding to the selected grid points in $\Theta$.

The procedure to synthesize the controller $K$, the assessment of closed-loop internal stability and the selection of a closed-loop performance cost function $J$ are only implicitly part of the grid point allocation problem, hence we formulate our approach under a freedom of choice for the user to choose the most appropriate objectives for a particular engineering application. As this paper focuses on the grid point allocation problem for robust and LPV analysis and control, we define appropriate synthesis procedures, stability conditions and closed-loop performance metrics for each case, see Subsections~\ref{sec:LPVapplication} and~\ref{sec:Robustapplication}, respectively.

\section{Methodology}\label{Sec:method}
Our main objective is to find the selection $\Theta$ for which a grid-based controller synthesis procedure can be employed to achieve the desired stability and closed-loop performance specifications globally.
To find the optimal selection $\Theta$, a systematic approach that aims at finding the most informative point within the admissible set regarding the desired closed-loop performance specifications is proposed. 

To propose an approach capable of accomplishing this selection process, we will first introduce an informativity notion for grid points on which we can establish an optimal  
%In the following subsections, a grid point informativity notion and the 
grid point allocation algorithm.
%are first proposed, and we 
Then, we introduce a Bayesian learning approach to make the grid point allocation computationally efficient. Next, we discuss how to use the grid point allocation approach specifically for robust and LPV controller design. Lastly, initialization strategies based on practical considerations and the available computational budget are discussed.

\subsection{Selection optimality and grid point informativity}
We define the selection $\Theta$ to be an \emph{optimal selection} w.r.t.~closed-loop stability and given performance specifications as follows.
\begin{definition}[$N_\theta$-optimal selection] \label{def:gridOptimality}
    Consider a given plant $P$ of the form~\eqref{eq:genplantRobust} or~\eqref{eq:LPV:op}, a given cost function $J$, a given set $\Theta \subseteq \Xi$ containing $N_\theta$ distinct points, i.e., $|\Theta|=N_\theta$, and a given controller $K$ synthesized via~\eqref{eq:grid_synth} with $P_{\Theta}$. Then, $\Theta$ is an $N_\theta$-optimal selection if for any set
    % $\Psi = \{\psi^{(i)}\}_i^{N} \subset \Xi$
    $\Psi \subseteq \Xi$ with $|\Psi| = N_\theta$ 
    and $\Psi \neq \Theta$, the following condition holds:
    \begin{equation}\label{eq:optimSetCond}
        J(P_\Theta, K)  \geq J(P_\Psi, K).
    \end{equation}
    %\TR{and $P_\theta \star K_\theta$ is internally stable for or all $\Xi$.}
\end{definition}
Then, to construct the optimal selection $\Theta$ with a prescribed cardinality, each newly added point must be \emph{the most informative} with respect to the closed-loop performance specifications. A point $\theta^\ast$ is regarded as the most informative if its inclusion, $\Theta \leftarrow \Theta \cup \{\theta^\ast\}$, results in the greatest degradation of the closed-loop performance. This is because the synthesis procedure~\eqref{eq:grid_synth} relies on the finite set of local models $P_\Theta$, and therefore cannot account for dynamic information not represented in $P_\Theta$. By identifying and including such a point $\theta^\ast$, the synthesis procedure can be updated to explicitly account for this critical dynamic information that was previously missing. The notion of the most informative grid point is formalized in the following definition.

\begin{definition}[Most informative grid point w.r.t.~$\Theta$] \label{def:informativity}
    Consider a given plant $P$ of the form~\eqref{eq:genplantRobust} or~\eqref{eq:LPV:op}, a given cost function $J$, given set $\Theta \subseteq \Xi$ containing $N_\theta$ distinct points and a given controller $K$ synthesized via~\eqref{eq:grid_synth} with $P_{\Theta}$. Then, $\theta^\ast$ is the most informative point w.r.t.~a selection $\Theta$, if
    \begin{equation} \label{eq:mostInformativeOptim}
        \theta^\ast \in \arg \max_{\theta} \ J(P_\theta, K) \quad \text{subject to} \quad \theta \in \Xi \setminus \Theta.
    \end{equation}
\end{definition}

It is important to note that similar to greedy optimization methods, it is generally not true that, for an $N_\theta$-optimal selection $\Theta$, the most informative point $\theta^\ast$ will satisfy that $\Theta \cup \theta^\ast$ is an $N_\theta+1$ optimal selection. However, such a construction can be arbitrary close to the optimal set, while requiring the solution of an optimization problem \eqref{eq:mostInformativeOptim} with limited complexity as $N_\theta$ grows.

\subsection{Systematic grid point allocation algorithm\label{sec:mainAlgorithm}}
We propose a three-step iterative grid point allocation algorithm to find a grid-point selection in terms of the optimality notion of Definition~\ref{def:informativity}. For the initialization of the algorithm, a generalized plant $P$, a cost function $J$, an initial selection $\Theta_0$ and an initial stabilizing controller $K_0$ synthesized based on $P_{\Theta_0}$ are required. Then, the iterative procedure is executed until the selection with prescribed cardinality $|\Theta| = N_\theta$ is obtained, as detailed in Algorithm~\ref{generalAlgorithm}. Alternatively, the iterative process can be executed until~\eqref{eq:mostInformativeOptim} fails to find a higher value of $J$, meaning that the allocated points in $\Theta$ account for all critical dynamic information w.r.t. the closed-loop performance objectives represented by $J$.
\begin{algorithm}[b]
    \caption{Grid point allocation algorithm} \label{generalAlgorithm}
    \textbf{Inputs:} Generalized plant $P$, cost function $J$, initial selection $\Theta_0$, initial stabilizing controller $K_0$, number of points to allocate $N_\theta$.\\
    \textbf{Outputs:} Allocated selection $\Theta$.
    \begin{algorithmic}[1]
        \State {\bf set} $\Theta \leftarrow \Theta_0$, $n \leftarrow  |\Theta_0|$, and $K \leftarrow K_0$.
        \While{$n< N_\theta$}
        \State {\bf solve}~\eqref{eq:mostInformativeOptim} to find a most informative point $\theta^\ast$.
        \State {\bf update} $\Theta \leftarrow \Theta \cup \theta^\ast$.
        \State {\bf synthesize} a controller $K$ for $P_\Theta$.
        \State {\bf set} $n \leftarrow n+1.$
        \EndWhile
        \State {\bf return} the grid-point selection $\Theta$.
    \end{algorithmic}
\end{algorithm}

However, a direct solution of~\eqref{eq:mostInformativeOptim} is computationally intractable in practice. This arises from typical performance metrics employed in existing controller synthesis methods, commonly based on induced $\mathcal{L}_{\mathrm p}$-$\mathcal{L}_{\mathrm q}$-gain, $\mathcal{H}_\infty$, $\mathcal{H}_2$, or combinations of these. Since $J$ is defined according to these metrics, problem~\eqref{eq:mostInformativeOptim} inherits their nonlinearity and non-smoothness, and each evaluation of $J$ entails a significant computational burden. 
To circumvent this challenge, we employ a Bayesian optimization approach to learn a surrogate cost function $\hat{J}$ that predicts the value of $J$ for unseen samples of $\theta$, while being inexpensive to evaluate and optimize. Then, by maximizing $\hat{J}$, the most informative grid point $\theta^\ast$ can be predicted in a computationally efficient manner. The following section details how Bayesian optimization is used to learn and optimize $\hat{J}$.

\subsection{Learning and global optimization of the cost function\label{GPandBO}}
% As direct optimization of the cost function $J$ defined in~\eqref{eq:mainobjective} is computationally intractable, we leverage the numerical efficiency of machine learning methods. In particular, we propose a Bayesian optimization approach to predict $\theta^\ast$.
Based on a set of observations of $J$, the Bayesian optimization approach begins by learning a surrogate cost surface $\hat{J}$ that predicts the value of $J$ at unseen samples. The surrogate cost $\hat{J}$ is learnt via a \emph{Gaussian process} (GP) regression model $\pazocal{GP} :\Real^{n_\mathrm{q}}\rightarrow \Real$. The uncertainty of the prediction of $J$ and the actual shape of $\hat{J}$ are blended in a scalar-valued function named as \emph{acquisition function}, 
% which is optimized to determine the next observation point based on which the most information about the global maximum of $J$ can be obtained. 
which is used to predict the next observation point with the highest information gain about the global maximum of $J$, and whose evaluations are computationally cheap.
At any step, $\hat{J}$ can be maximized to predict the global maximum of $J$. 
This process can be performed iteratively, where for each iteration a new observation point is determined, the true cost function $J$ is evaluated, and the surrogate cost surface $\hat{J}$ is updated. This process stops when it reaches a user-defined number of iterations, which can be chosen based on the available computational budget.

% Therefore, for a given set $\Theta_N = \{\theta^{(i)}\}_i^N$, a given generalized plant $P$ of the form of~\eqref{NLgen}, a given stabilizing controller $K$ and a given set of observations $\mathbb{D}_N \coloneq \{ \gamma_i = J(P_{\mathrm{LTI}}^{(\theta, \; i)}, K), \ \theta^{(i)} \}_{i=1}^{N}$, the first step is to learn a GP regression model that approximates the relation ${J}:(P_{\mathrm{LTI}}^{(\theta)}, K) \mapsto \gamma$ between the input $\theta \in \Xi$ and output observations $\gamma \in \Real$ of the form
Therefore, for a given set of observations $\mathbb{D}_N \coloneq \{ \gamma_i = J(P_{\theta^{(i)}}, K), \ \theta^{(i)} \}_{i=1}^{N}$, the first step is to learn a GP regression model that approximates the relation ${J}:(P_\theta, K) \mapsto \gamma$ between the input $\theta \in \Xi$ and output observations $\gamma \in \Real$ of the form
\begin{align} \label{eq:datagen_relation}
    \gamma = \hat{J}(P_\theta, K).
\end{align}
The main idea of GP-based estimation of $J$ is to consider that candidate estimates $\hat{J}$ belong to a GP, seen as \emph{prior} distribution. With this prior and $\mathbb{D}_N$, a \emph{posterior} GP distribution of $\hat{J}$ that provides estimates of $J$ by its mean and describes uncertainty of this estimate by its variance is computed\cite{rasmussenGaussianProcessesMachine2006a}. 

The first step is to assign a random variable $\pazocal{GP}(\theta)$ to every point $\theta$, such that for any finite set $\theta^{(1)}, \dots, \theta^{(N)}$, the joint probability distribution of $\pazocal{GP}(\theta^{(1)}), \dots, \pazocal{GP}(\theta^{(N)})$ is Gaussian. 
% A GP $\pazocal{GP} :\Real^{\nx} \times \Real^{\dnu} \times \Real^{n_\mathrm{q}}\rightarrow \Real$ assigns to every point $\theta$ a random variable $\pazocal{GP}(\theta)$, such that for any finite set $\theta^{(1)}, \dots, \theta^{(N)}$, the joint probability distribution of $\pazocal{GP}(\theta^{(1)}), \dots, \pazocal{GP}(\theta^{(N)})$ is Gaussian. 
Then, $\pazocal{GP}$ is fully characterized by its mean $m$ and its covariance $\kappa$ functions, such that if $\hat{J} \sim \pazocal{GP}(m, \kappa)$, where
\begin{equation}
    m(\theta) = \mathbb{E}\{\hat{J}(\theta)\}, \qquad \kappa(\theta, \tilde{\theta}) = \mathbb{E} \{\left(\hat{J}(\theta) - m(\theta) \right) \left(\hat{J}(\tilde{\theta}) - m(\tilde{\theta}) \right) \},
\end{equation}
then the joint Gaussian probability of $\pazocal{GP}(\theta^{(1)}), \dots, \pazocal{GP}(\theta^{(N)})$ is described by the normal distribution $\mathcal{N}(M_\Theta,K_{\Theta \Theta})$, with
\begin{equation}
    M_\Theta =\left(\ m\left(\theta^{(1)}\right)\ \cdots\ m\left(\theta^{(N)}\right)\ \right)^{\top}, \quad  [K_{\Theta \Theta}]_{i,j} =\kappa\left(\theta^{(i)}, \theta^{(j)}\right)_{i, \, j = 1}^{N}.
\end{equation}
Both $m$ and $\kappa$, where $\kappa$ is also named as \emph{kernel} function due to the relation of GPs to \emph{reproducing kernel Hilbert space} estimators\cite{rasmussenGaussianProcessesMachine2006a},
are parametrized in terms of the \emph{hyperparameters} $\lambda \in\mathbb{R}^{n_\lambda}$, which allows to adjust the prior to the approximation of $J$ using $\mathbb{D}_N$. Although many kernel functions $\kappa$ have been introduced in the machine learning literature for GP regression \cite{rasmussenGaussianProcessesMachine2006a}, a widely used choice in the Bayesian optimization context is the \emph{mat\'{e}rn 5/2} ($C_{5/2}$) kernel, given by
\begin{equation}\label{kernelc52}
    \kappa_\mathrm{C_{5/2}}(\theta, \tilde{\theta}) = \lambda_1^2 \left(1 + \frac{\sqrt{5}d}{\lambda_2} + \frac{5 d^2}{3 \lambda_2^2} \right) \exp\left(- \frac{\sqrt{5}d}{\lambda_2}\right),
\end{equation}
where $d$ is the distance between $\theta$ and $\tilde{\theta}$, typically taken as the euclidean distance, i.e. $d^2 = \left(\theta - \tilde{\theta}\right)^\top \left(\theta - \tilde{\theta}\right)$.

Then, based on $\mathbb{D}_N$ and the prior $\hat{J} \sim \pazocal{GP}$, %the prior given by
\begin{equation}\label{eq:prior}
    \mathrm{pr}(\, \Gamma \, | \,  \Theta, \lambda) = \mathcal{N}(M_\Theta, K_{\Theta \Theta}),
\end{equation}
describes the probability density function of the observed outputs $\Gamma = ( \gamma^{(1)} \ \cdots \ \gamma^{(N)})^\top$, where $\Gamma$ are seen as random variables conditioned on the inputs $\Theta = \{ \theta^{(1)},\ \ldots, \ \theta^{(N)}\}$ and the hyperparameter values $\lambda$. In our case, we can assume that the prior mean is the zero function $m(\theta) = 0$ if no reliable prior information about $J$ is available, while the choice of $\kappa$ determines the function space in which an estimate of $J$ is searched for. The joint distribution
\begin{align*}
    \begin{pmatrix}
        {\Gamma}             \\
        \hat{J}(\theta^\ast) \\
    \end{pmatrix}
    \sim \mathcal{N}\left( \begin{pmatrix}
                                   M_\Theta \\
                                   m(\theta^\ast)
                               \end{pmatrix}, \begin{pmatrix}
                                                  K_{\Theta \Theta} & K_{\Theta}(\theta^\ast) \\ K_{\Theta}^\top (\theta^\ast)&   \kappa(\theta^\ast,\theta^\ast)
                                              \end{pmatrix}
    \right)
\end{align*}
with $[ K_\Theta(\theta^\ast) ]_i =\kappa(\theta^{(i)},\theta^\ast)$ can be used to estimate the value of the original function $J$ at a query point $\theta^\ast$. The predictive distribution that estimates $J(\theta^\ast)$ based on $\mathbb{D}_N$ is the posterior distribution $\mathrm{pr}\!(\hat{J}(\theta^\ast) \, | \, \mathbb{D}_N, \theta^\ast) = \mathcal{N}(\mu(\theta^\ast),\sigma(\theta^\ast))$, where
\begin{subequations}\label{eq:posterior}
    \begin{align}
        \mu(\theta^\ast)    & = m(\theta^\ast) +K_\Theta^\top(\theta^\ast) \, K_{\Theta \Theta}^{-1} \, \left(\Gamma - M_\Theta \right), \label{eq:posterior_m} \\
        \sigma(\theta^\ast) & =
        k(\theta^\ast, \theta^\ast)-{K_\Theta^\top (\theta^\ast)} \, K_{\Theta \Theta}^{-1} \, K_\Theta (\theta^\ast). \label{eq:posterior_s}
        %    
        %    
        %        \begin{split}\label{eq:posterior_m}
        %            \mu(x_*) &= \mathbb{E}[y\mid [X, x_*], \bar y, \theta] \\&= m(x_*) + \kappa(x_*, X)(K+ \sigma_\epsilon^2I)^{-1}(\bar y-m(X)),
        %        \end{split}\\
        %        \begin{split}\label{eq:posterior_s}
        %        \sigma(x_*) &= \mathrm{var}[y\mid [X, x_*],\bar y, \theta] \\&=  \kappa(x_*,x_*) -  \kappa(x_*, X)(K + \sigma_\epsilon^2I)^{-1}\kappa(X, x_*),
        %        \end{split}
    \end{align}
\end{subequations}

The hyperparameters $\lambda$ associated with the prior distribution represent the degrees of freedom of the learning prcoess. A common strategy \cite{bishopPatternRecognitionMachine2016} to find the best hyperparameter values to represent the observed data $\mathbb{D}_N$ is to maximize the logarithm of the marginal likelihood, which represents the probability of generating the observations $\Gamma$ from the prior distribution \eqref{eq:prior} marginalized with respect to $\lambda$:
\begin{align} \label{hyp:opt}
    \lambda^* = \arg \max_{\lambda} \ \log \left(\mathrm{pr}\!\left(\, \Gamma \, | \, \Theta, \lambda\right) \right),
    %\nonumber
\end{align}
where
% \vspace{-2mm}
\begin{equation}
    \log \left(\mathrm{pr}\!\left(\, \Gamma \, | \, \Theta, \lambda \right) \right)  = -\frac{1}{2} \left( \Gamma^\top  K_{\Theta \Theta}^{-1} \, \Gamma + \log \det   K_{\Theta \Theta}^{-1} + N\log 2\pi \right).
\end{equation}
is the logarithm of the probability distribution function $\mathrm{pr}\!\left(\, \Gamma \, | \, \Theta, \lambda\right)$ that is taken to simplify the optimization problem.

Once the GP regression model $\hat{J}$ that estimates $J$ is learned, the second step in BO is to construct the so called acquisition function. A well known acquisition function is the \emph{expected improvement} (EI), which is defined as follows. 
Suppose that, for a given set of observations $\mathbb{D}_N$, a given generalized plant $P$ and a given stabilizing controller $K$, the predictive distribution $\hat{J}_N \sim \pazocal{GP}\left(m_N, \kappa_N \right)$ is learned, where $m_N$ and $\kappa_N$ are defined as in~\eqref{eq:posterior}. 
Let $\gamma^+$ be the maximum observed value of $J$ that has been encountered so far, i.e., $\gamma^+ = \max J(P_{\Theta_N}, K)$. Then, the EI is defined as
\begin{equation} \label{expectedimprovement}
    \textrm{EI}_N \coloneq \mathbb{E} \left( \max\{0, \ \hat{J}(P_\theta, K) \} - \gamma^+ | \ \mathbb{D}_N\right),
\end{equation}
where the right-hand-side of \eqref{expectedimprovement} has the analytic form
\begin{align}
    \textrm{EI}_N(\theta, \varepsilon) = \begin{cases}
                                             \left(\mu_N(\theta) - \gamma^+ - \varepsilon  \right) \Phi(Z) + \sigma_N(\theta) \, \phi(Z), & \textrm{if} \quad  \sigma_N(\theta)>0; \\
                                             0,                                                                                           & \textrm{if} \quad  \sigma_N(\theta)=0;
                                         \end{cases}
     & \quad
    Z = \frac{\mu_N(\theta) - \gamma^+ - \varepsilon}{\sigma_N(\theta)},
\end{align}
with $\Phi(\cdot)$ being the cumulative distribution function, $\phi(\cdot)$ the probability density function and $0< \varepsilon \in \Real$ is a constant typically known as the \emph{exploration ratio}. Other well-known acquisition functions in the literature are the upper confidence bound or probability of improvement \cite{Frazier:2018iqo}. Then, the next point to observe with the highest expected information gain about the global maximum of $J$ is obtained by solving the optimization problem
\begin{equation}\label{acqfunopt}
    \hat{\theta} \in \arg \max_{\theta \in \Xi} \textrm{EI}_N(\theta, \varepsilon).
\end{equation}
Unlike the true objective function $J$, evaluating $\mathrm{EI}_N(\theta, \varepsilon)$ is computationally cheap, and there exist a variety of approaches to solve~\eqref{acqfunopt} in a computationally efficient manner, such as greedy or gradient-based methods\cite{brochu2010, wilson2018maximizing}.

\begin{algorithm}[t]
    \caption{Bayesian optimization algorithm to find a most informative grid point $\theta^\ast$}
    \label{alg:bayesOpt}
    \textbf{Inputs:} Generalized plant $P$, cost function $J$, observation data set $\mathbb{D}_N = \{ \gamma_i = J(P_{\theta^{(i)}}, K), \ \theta^{(i)}\}_{i=1}^{N}$, stabilizing controller $K$, prior mean function $m$, kernel function $\kappa$, exploration ratio $\varepsilon$, maximum number of iterations $N_\mathrm{max} \in \mathbb{Z^+}$.\\
    \textbf{Outputs:} Predicted most informative grid point $\theta^\ast$.
    \begin{algorithmic}[1]
        \While {$N < N_\mathrm{max}$}
        \State {\bf compute} $\hat{J}_N \sim \pazocal{GP}(\mu_N, \sigma_N)$ based on $\mathbb{D}_N$ via \eqref{eq:posterior}.
        \State {\bf solve} \eqref{hyp:opt} for $\hat{J}_N$ to find the hyperparameters $\lambda$ %with of $\hat{J}_N$ via \eqref{hyp:opt}.
        \State {\bf set} $\hat{\theta} \in \arg \max_{\theta \in \Xi} \mathrm{EI}_N(\theta, \varepsilon)$.
        \State {\bf observe} $\hat{\gamma} \leftarrow J(P_{\hat{\theta}}, K)$.
        \State {\bf set} $\mathbb{D}_{N+1} \leftarrow \mathbb{D}_N \cup (\hat{\gamma}, \hat{\theta})$ and $N\leftarrow N+1$.
        \EndWhile
        \State  {\bf return} $\theta^{\ast} \in \arg \max_{\theta \in \Xi} \mu_N(\theta)$.
        % \State  Return $\theta^{\ast} \leftarrow \max\{\arg \max_{\theta \in \Xi} \mu_N(\theta), \ \theta \in \mathbb{D}_N\}$.
    \end{algorithmic}
\end{algorithm}
The described Bayesian optimization procedure can be used to approach the optimization problem~\eqref{eq:mostInformativeOptim} in Algorithm~\ref{generalAlgorithm} and predict the most informative point $\theta^\ast$, as summarized in Algorithm~\ref{alg:bayesOpt}. 

\subsection{Application for gridded LPV controller design\label{sec:LPVapplication}}
As indicated in Section~\ref{Sec:Problem_Statement}, the proposed approach is flexible enough to accommodate various choices for the assessment of closed-loop internal stability and performance together with the used controller synthesis procedure. In this subsection, we elaborate on how to tailor the grid point allocation method specifically for LPV controller design for which we consider generalized plants in the form of~\eqref{eq:LPV:op}.

To assess the closed-loop internal stability, we rely on the point-wise form of~\eqref{eq:LPVinternalstability} with a constant $X$, i.e., $P_\theta \star K_\theta$ is internally stable on $\Theta$ if there exists a matrix $X \in \Real^{(n_{\mathrm{x} }+n_{\mathrm{x_k}}) \times (n_{\mathrm{x} }+n_{\mathrm{x_k}})}$ such that
\begin{equation}\label{eq:LPVinternalstability_pointwise}
{A}_{\mathrm{cl}}^\top(\theta) X + X {A}_{\mathrm{cl}}(\theta) \prec 0, \quad X \succ 0, \quad \forall \theta\in\Theta.
\end{equation}
Note that this test does not provide guarantees for the global closed-loop stability of the entire LPV system, until $\{{A}_{\mathrm{cl}}(\theta)\}_{\theta\in\Xi}\subseteq \mathrm{conv}( \{{A}_{\mathrm{cl}}(\theta)\}_{\theta\in\Theta})$ with $ \mathrm{conv}(\bullet)$ being the convex hull. Hence in general, stability holds  only locally at the grid points. In terms of the closed-loop performance metric, consider the point-wise induced $\mathcal{L}_p$-$\mathcal{L}_q$-gain cost:
\begin{equation} \label{eq:LpLqGain}
    J(P_\theta, K_\theta) \coloneq \sup_{0 < \norm{w}_p < \infty} \frac{\norm{P_\theta \star K_\theta \ w}_{q}}{\norm{w}_{p}} \leq \gamma, \quad  w \in \pazocal{W}, \ \theta \in \Xi,
\end{equation}
where $\norm{\bullet}_{p}$ and $\norm{\bullet}_{q}$ are the $\mathcal{L}_p$ and $\mathcal{L}_q$ norm of the signals $w$ and $P_\theta \star K_\theta \ w$, respectively.
% %
% \begin{equation} \label{eq:LpLqGain}
%     J(P_\theta, K_\theta) \coloneq \operatorname{ess} \sup \frac{\norm{P_\theta \star K_\theta \ w}_{q,T}}{\norm{w}_{p,T}} \leq \gamma, \quad \theta \in \Xi, \ \forall \ T \geq 0, \, \norm{w}_{p,T} \neq 0, \, w \in \pazocal{W},
% \end{equation}
% %
% where $\norm{\cdot}_{p,T}$ and $\norm{\cdot}_{q,T}$ are the truncated $p$-norm and $q$-norm of the signals $w$ and $P_\theta \star K_\theta \ w$, respectively, and $\mathcal{L}_{\mathrm{pe}}$ denotes the extended $\mathcal{L}_{\mathrm{p}}$ space, adopted from \cite{vanderschaftL2GainPassivityTechniques2016}.
If $p=q$ is chosen, we will refer to this as the induced $\mathcal{L}_p$-gain. One of the most used performance metrics based on the induced $\mathcal{L}_p$-$\mathcal{L}_q$-gain is the induced $\mathcal{L}_2$-gain, which can be understood as the bound on the amplification of the input signal $w$ with finite energy to the output of the closed-loop system $P_\theta \star K_\theta$.
Finally, there exist several grid-based LPV controller synthesis methods in the literature, see~\cite{hoffmannSurveyLinearParameterVarying2015a} for an overview. In this work, we consider the methods based on Corollaries 2.5, 2.7 and 2.8 in~\cite{koelewijnAnalysisControlNonlinear2023}, as these correspond to the $\mathcal{L}_2$-gain, the $\mathcal{L}_2$-$\mathcal{L}_\infty$-gain and the $\mathcal{L}_\infty$-gain, respectively, and are readily implemented in the \textsc{LPVcore} toolbox for \textsc{Matlab}\cite{DENBOEF2021385}. However, these gridded synthesis methods produce a set of local LTI controllers at the (scattered) points in $\Theta$, i.e., $K_\Theta \coloneq \{K^{(i)}\}_i^{N_\theta}$, where
\begin{equation}\label{eq:LPVcontrollersnapshot}
    K^{(i)} \coloneq \begin{cases}
        \dot{x}_\mathrm{k} = \mathtt{A}_\mathrm{k}^{(i)} x_\mathrm{k} + \mathtt{B}_\mathrm{k}^{(i)} u_\mathrm{k}, \\
        y_\mathrm{k} = \mathtt{C}_\mathrm{k}^{(i)} x_\mathrm{k} + \mathtt{D}_\mathrm{k}^{(i)} u_\mathrm{k}.
    \end{cases}
\end{equation}
To realize the complete LPV controller $K_p$ of the form~\eqref{eq:LPVcontroller} from $K_\Theta$, an $\np$-dimensional interpolation scheme is required to retrieve the controller matrices at any point of the admissible scheduling signal space. In practise, if Algorithm~\ref{generalAlgorithm} is employed to find $\Theta$ for synthesizing $K_\Theta$, the resulting $\Theta$ cannot be assumed to form a regular grid\footnote{A regular grid consists of points that are uniformly spaced over each dimension}. Consequently, a multidimensional \emph{scattered} interpolation scheme is required to construct $K_p$. Inspired by~\cite{amidrorScatteredDataInterpolation2002}, we propose the following \emph{radial basis functions} (RBF)-based scattered interpolation scheme. For this, we first construct the matrix $M_\Theta \in \Real^{N_\theta \times (n_x + n_y)\cdot(n_x + n_u)}$ given by
\begin{equation}
    M_\Theta  \coloneq \begin{pmatrix}
        \operatorname{vec}(M^{(1)})^\top & \cdots\ & \operatorname{vec}(M^{(N_\theta)})^\top
    \end{pmatrix}^\top
    \qquad \text{with} \qquad M^{(i)}  = \begin{pmatrix}
        \mathtt{A}_k^{(i)} & \mathtt{B}_k^{(i)} \\ \mathtt{C}_k^{(i)} & \mathtt{D}_k^{(i)}
    \end{pmatrix},
\end{equation}
where $\operatorname{vec}(\bullet)$ denotes the row-wise vectorization of a matrix. In addition, let $\operatorname{mat}(\bullet)$ denote the inverse operation, i.e., $\operatorname{mat}(\operatorname{vec}(M)) = M$. Then, consider the multiquadratic RBF distance function $d : \Real^{\np \times 2} \rightarrow \Real$ given by
\begin{equation} \label{eq:RBFmultiquadric}
    d(\theta, \tilde{\theta}) \coloneq \sqrt{\|\theta - \tilde{\theta} \|_2^2 + c^2},
\end{equation}
where $c > 0, \, c \in \Real$ is a constant. Next, we construct the weighting matrix \mbox{$W \in \Real^{N_\theta \times (n_x + n_y)\cdot(n_x + n_u)}$} as
\begin{equation} \label{eq:RBFweight}
    W = D_{\Theta \Theta}^{-1} M_\Theta, \qquad \text{with} \qquad D_{\Theta \Theta} \coloneq d(\theta^{(i)}, \theta^{(j)}) \quad \text{for} \quad i, j \in \idx{1}{N_\theta}, \quad D_{\Theta \Theta} \in \Real^{N_\theta \times N_\theta}.
\end{equation}
Moreover, let us denote the rows of $W$ in~\eqref{eq:RBFweight} as
$W_i \in \Real^{1 \times (\nx+\ny)\cdot(\nx+\dnu)}$. Now it holds that
\begin{equation}
    M_i = \operatorname{mat}\left( \sum_{j=1}^{N_\theta} d(\theta^{(i)}, \theta^{(j)}) W_j \right) \quad \text{for} \quad i \in \idx{1}{N_\theta},
\end{equation}
which shows that the local controller matrices $M_i$ can be reconstructed without approximation error at any point $\theta \in \Theta$. Furthermore, to query an approximation of the controller matrices at any point $p \in \mathcal{P}$, we take
\begin{equation}\label{eq:RBFquery}
    \hat{M}(p) = \operatorname{mat}\left( \sum_{j=1}^{N_\theta} d(p, \theta^{(j)}) W_j \right).
\end{equation}
Then, based on~\eqref{eq:RBFquery}, a realization of the controller $K$ in~\eqref{eq:LPVcontroller} at any point $p \in \mathcal{P}$ is given by
\begin{equation}\label{eq:RBFcontroller}
    \begin{pmatrix}
        \dot{x}_k(t) \\ y_k(t)
    \end{pmatrix} = \hat{M}(p(t)) \begin{pmatrix}
        x_k(t) \\ u_k(t)
    \end{pmatrix}.
\end{equation}

\subsection{Application for gridded robust controller design\label{sec:Robustapplication}}
In this subsection, following a similar structure as in the LPV counterpart, we elaborate provide a tailored grid point allocation method for gridded robust controller design for which we consider generalized plants in the form of~\eqref{eq:genplantRobust}.

To assess the closed-loop internal stability, we use the transfer matrix test in~\eqref{eq:9blocktest} evaluated at the grid locations, or its alternative SS form, i.e., $P_\theta \star K$ is internally stable at $\theta \in \Theta$ if:
\begin{equation*}
    \rho(A_{\mathrm{cl}}(\theta)) < 0 \quad \text{for all} \quad \theta \in \Theta,
\end{equation*}
where $\rho(\bullet)$ denotes the real spectral radius. 
For closed-loop performance,  robust performance metrics, given by
\begin{equation} \label{eq:RPcost}
    J(P_\theta, K) \coloneq \|\theta \star P \star K \|, \quad \theta \in \Theta,
\end{equation}
are used, 
where $\| \bullet \|$ denotes any system norm, e.g., the $\mathcal{H}_\infty$ or the $\mathcal{H}_2$ norm. Moreover, note that $J$ in~\eqref{eq:RPcost} also allows the use of different system norms for different input-output pairs, i.e.,
\begin{equation} \label{eq:RPcostmulti}
    J(P_\theta, K) \coloneq \max\left\{\|T_{w_i \to z_i}(\theta, K) \|_2, \ \|T_{w_j \to z_j}(\theta, K) \|_\infty \right\},
\end{equation}
where $T_{w_{(\bullet)} \to z_{(\bullet)}}(\theta, K)$ denotes the closed-loop map
$\theta \star P \star K$ from signals $w_{(\bullet)}$ to signals $z_{(\bullet)}$. Lastly, the controller synthesis becomes a multi-model (and possibly multi-objective) problem, for which we use the results from~\cite{apkarianNonsmoothSynthesis2006, apkarianMultimodelMultiobjectiveTuning2014} for fixed-structure controller synthesis, which are implemented in  \textsc{systune}~\cite{RobustControlToolbox} in \textsc{Matlab}.
% Finally, in terms of controller synthesis, for a given grid set $\Theta$ containing $N_\theta$ realizations of the uncertainty, the goal is to find a controller $K$ of the form~\eqref{eq:robustController} that stabilizes $P_{\mathrm{LTI}}^{(\theta)}$ while minimizes $J$ in~\eqref{eq:RPcostmulti} for all $\theta \in \Theta$. This can be formulated as the following multimodel, multi-objective controller synthesis problem:
% %
% \begin{equation}\label{eq:synthesisRobust}
%     K^\ast = \arg\min_{K} J(P_{\mathrm{LTI}}^{(\theta)}, K) \quad \text{such that} \quad K \text{ stabilizes } P_{\mathrm{LTI}}^{(\theta)} \quad \text{for all} \quad \forall \, \theta \in \Theta.
% \end{equation}
% %
% To solve this synthesis problem, we use the results from~\cite{apkarianNonsmoothSynthesis2006, apkarianMultimodelMultiobjectiveTuning2014} for fixed-structure controller synthesis, which are implemented in the \textsc{systune} solver~\cite{RobustControlToolbox} from \textsc{Matlab}.

\subsection{Practical considerations of the grid-point allocation algorithm\label{sec:initialization}}
In either the LPV or the robust case, the grid point allocation Algorithm~\ref{generalAlgorithm} requires an initial selection $\Theta_0$ and a stabilizing controller $K_0$ for its initialization. In principle, any suitable strategy can be employed for this purpose, till the resulting initial $K_0$ design based on $\Theta_0$ internally stabilizes the plant on the whole $\Xi$.
For instance, $\Theta_0$ can be composed of the extrema / vertices of $\Xi$, and/or by including points of particular dynamic interest. 
However, in practice, finding a $K_0$ through the selection of $\Theta_0$ that internally stabilizes the entire plant is not always straightforward, as the same issues to those discussed in Section~\ref{sec:intro} can be encountered, even when the closed-loop performance objectives are not considered.

Hence, the suggested approach is to employ a random initialization procedure in which $\Theta_0$ is constructed by sampling $N$ randomly distributed points contained in $\Xi$, followed by the synthesis of an initial controller $K_0$ that locally stabilizes the plant on $\Theta_0$, as described in Algorithm~\ref{alg:randomInit}. Although the resulting $K_0$ does not guarantee global stability for all $\theta \in \Xi$, in practise the global stability requirement in Algorithm~\ref{generalAlgorithm} can be relaxed to a local stability requirement. This relaxation is justified because the performance metrics~\eqref{eq:LpLqGain} and~\eqref{eq:RPcost} are finite only when the closed-loop system is internally stable, and any $\theta$ for which $K_0$ fails to stabilize the plant yields an unbounded value of the cost $J(\theta)$. Consequently, maximization in~\eqref{eq:mostInformativeOptim} naturally guides the search towards unstable regions, ensuring that any $\theta \in \Xi$ not stabilized by $K_0$ is detected. However, it must be emphasized that this argument holds only if Algorithm~\ref{alg:bayesOpt} succeeds in finding the global optimum of~\eqref{eq:mostInformativeOptim}.

The most computationally expensive operations of the proposed method are the computation of the local models $P_\theta$ and evaluations of the cost function $J(P_\theta, K_\theta)$. The parameters of Algorithm~\ref{alg:bayesOpt} allow explicit control over the computational complexity of the grid point allocation procedure. Although most of these parameters follow naturally from Algorithm~\ref{generalAlgorithm}, the construction of the observation set $\mathbb{D}_N$ involves a trade-off between the likelihood of finding the most informative grid point and the computational cost. The set $\mathbb{D}_N$ is used to train the GP regression model and increasing its cardinality $N$ can improve the accuracy of the GP regression model that estimates $J$. Moreover, $\mathbb{D}_N$ is not only restricted to the points $\theta \in \Theta$ generated with Algorithm~\ref{generalAlgorithm} as also different points can be employed to increase the exploration of the admissible space.
Similarly, increasing the number of iterations $N_{\mathrm{max}}$ in Algorithm~\ref{alg:bayesOpt} directly enhances the probability of identifying the global optimum of~\eqref{eq:mostInformativeOptim} at the expense of one additional local model request and cost function evaluation per iteration. Hence, estimates of the computational cost required for obtaining a local model and for evaluating the cost function can be employed to tune Algorithm~\ref{alg:bayesOpt} according to the available computational budget.
\begin{algorithm}[t]
    \caption{Initialization process}
    \label{alg:randomInit}
    \textbf{Inputs:} Required number of initial grid points $N_0$\\
    \textbf{Outputs:} Initial selection $\Theta_0$ and controller $K_0$.
    \begin{algorithmic}[1]
        \State {\bf draw} $N$ points $\Theta_0 = \{\theta^{(i)}\}_{i=1}^{N_0}$ randomly within the admissible set $\Xi$.
        \State {\bf synthesize} $K_0$ which locally stabilizes $P_{\Theta_0}$.
        \State {\bf return} $\Theta_0$ and $K_0$.
    \end{algorithmic}
\end{algorithm}
\section{Simulation Study}\label{Sec:simulation}
The proposed method is tested on the LPV and robust $\pazocal{H}_{\infty}$ grid-based controller design problem for
\begin{enumerate}
    \item A simple model that describes an unbalanced disk system with two uncertain parameters.
    \item A satellite that consists of a rigid body with two flexible rotating solar arrays and uncertain model parameters. The rotation of the solar arrays is represented with real parametric uncertainties.
    \item A two-degrees of freedom LPV model of a robotic arm with ten scheduling variables.
\end{enumerate}
The selection of these benchmarks is motivated as follows. The uncertain unbalanced disk provides a visually intuitive example to illustrate the proposed method. The satellite model with flexible rotating arrays is of industrial interest; as shown in \cite{sanfedinoExperimentalValidationHigh2019a}, the minimal LFR of a single rotating appendage involves one uncertainty parameter with sixteen repetitions, posing a significant computational challenge for current worst-case analysis tools. Finally, the LPV robotic arm model demonstrates the capabilities of the method for LPV controller design in high-dimensional scheduling space.
% With the uncertain unbalanced disk we intend to present the proposed method in a manner that is easy to visualize. The satellite model with flexible rotating arrays is of industrial interest, because as shown in \cite{sanfedinoExperimentalValidationHigh2019a}, the minimal \emph{linear fractional representation} (LFR) of a single rotating appendage requires one uncertainty parameters with sixteen repetitions, which is computationally challenging for the current worst-case analysis tools. Lastly, the LPV model of the robotic arm intends to show the capabilities of the proposed method for LPV controller design under a high-dimensional parameter space.
% With the uncertain unbalanced disk we intend to show the steps and capabilities of the proposed method in a manner that is easy to visualize. The satellite model with flexible rotating arrays is of industrial interest, because as shown in \cite{sanfedinoExperimentalValidationHigh2019a}, the minimal \emph{linear fractional representation} (LFR) of a single rotating appendage requires one uncertainty parameters with sixteen repetitions, which is computationally challenging for the current worst-case analysis tools. Lastly, the LPV model of the robotic arm with ten scheduling parameters intends to show the capabilities of the proposed method for LPV controller design under a high-dimensional parameter space.
\subsection{Unbalanced disk\label{sec:unbalanced_disk}}
In this example, we address a controller design problem for an uncertain model of an unbalanced disk. A robust grid-based structured control synthesis problem is first formulated to meet specific closed-loop performance requirements. Then, controllers are synthesized using four different grid sets, generated via: (i) an initial guess; (ii) equidistant gridding over the uncertainty space; (iii) a $\mu$-analysis-based grid allocation strategy; and (iv) the proposed grid point allocation method. Lastly, the resulting grid sets and the closed-loop performance of their respective controllers are subsequently compared. 

The unbalanced disk system consists of an actuated disc with a mass mounted on it and its motion can be described by
% , as shown in Fig.~\ref{fig:unbalanced_disk}.
%
% \begin{figure}[t]
%     \centering
%     \includegraphics[width=0.15\columnwidth]{figures/unbalanced_disk_figure.jpg}
%     \caption{Picture of the unbalanced disk system.}
%     \label{fig:unbalanced_disk}
% \end{figure}
%
% The motion of the unbalanced disk system can be described by
%
\begin{equation} \label{eq:unbalDiskNL}
    \Sigma \coloneq \left\{
    \begin{aligned}
        \dot{x}_1(t) & = x_2(t),                                      \\
        \dot{x}_2(t) & = -M c_1 \sin(x_1(t)) - c_2 x_2(t) + c_3 u(t), \\
        % \dot{x}_2(t) & = M c_1 \sin(x_1(t)) - \frac{1}{\tau}x_2(t) + \frac{K_m}{\tau}u(t) \\
        y(t)         & = x_1(t),
    \end{aligned}
    \right.
\end{equation}
where $x_1$ and $x_2$ represent the angle of the disk in radians and its angular velocity in radians per second, respectively, $u$ is the input voltage to the actuator in Volts, $y$ is the output of the system, $c_1 = {g l}/{J}$, $c_2 = {1}/{\tau}$, $c_3 = {K_m}/{\tau}$ and $M = \num{7e-2} \ \mathrm{kg}$, $g = \num{9.8} \ \mathrm{m \cdot s^{-2}}$, $l = \num{4.2e-2} \ \mathrm{m}$, $J = \num{2.2e-4} \ \mathrm{kg \cdot m^2}$, $\tau = \num{5.971e-1} \ \mathrm{s}$, and $K_\mathrm{m} = \num{1.531e1} \ \mathrm{rad \cdot s^{-1} \cdot V^{-1}}$ are the nominal values of the physical parameters. The weight of the off-centered mass $M$ is considered to be uncertain and it can have a variation of $\pm 60\%$ with respect its nominal value $M$. This uncertainty is modeled as ${M} = \hat{M} + w_M \delta_1$, where $w_M = \num{0.042}$, and $|\delta_1| < 1$ represents a real parametric uncertainty. Further, for this example, we approximate the nonlinearity in $\Sigma$ with another real parametric uncertainty. For this, we re-write the $\dot{x}_2$ dynamics of $\Sigma$ as
\begin{equation}
    \dot{x}_2(t) = -M c_1 \frac{\sin(x_1(t))}{x_1(t)}x_1(t) - c_2 x_2(t) + c_3 u(t),
\end{equation}
and define $p \approx \frac{\sin(x_1(t))}{x_1(t)} = \operatorname{sinc}(x_1(t))$, where $p \in [-0.22, 1]$. Then, let $p$ be described as $p = \hat{p} + w_p \delta_2$, where $\hat{p} = 0.39$ is the nominal value, $w_p = 0.61$ is the scaling weight and $|\delta_2| < 1$ represents a real parametric uncertainty. Based on these, $\Sigma$ is described with the LFR
\begin{subequations} \label{eq:LFR_unbalanced_disk}
    \begin{equation}
        G :=
        \begin{aligned}
            \left(\begin{array}{c}
                          \dot{x}_1 \\ \dot{x}_2 \\ \hdashline z_{\Delta, 1} \\ z_{\Delta, 2} \\ \hdashline y
                      \end{array}\right) & = \left(\arraycolsep=3pt \begin{array}{cc:ccc:c}
                                                                        0 & 1 & 0 & 0 & 0 & 0 \\ -c_1 \hat{M} \hat{p} & -c_2 & -c_1 w_M \hat{p} & -c_1 \hat{M} w_p & -c_1 w_M w_p & c_3 \\ \hdashline 1 & 0 & 0 & 0 & 0 & 0 \\ 0 & 0 & 1 & 0 & 0 & 0 \\ \hdashline 1 & 0 & 0 & 0 & 0 & 0
                                                                    \end{array} \right) \left(\begin{array}{c}
                                                                                                  x_1 \\ x_2 \\ \hdashline w_{\Delta, 1} \\ w_{\Delta, 2} \\ w_{\Delta, 3} \\\hdashline u
                                                                                              \end{array}\right)
        \end{aligned} ,
    \end{equation}
    where 
    \begin{subequations} %\label{eq:uncertainty_channels}
    \begin{align*}
        z_{\Delta, 1} & = x_1,                    & z_{\Delta, 2} & = w_{\Delta, 1},                                                      \\
        w_{\Delta, 1} & = \delta_1 z_{\Delta, 1}, & w_{\Delta, 2} & = \delta_2 z_{\Delta, 1}, & w_{\Delta, 3} & = \delta_2 z_{\Delta, 2},
    \end{align*}
\end{subequations}
    and
    \begin{equation}
        \begin{pmatrix}
            w_{\Delta, 1} & w_{\Delta, 2} & w_{\Delta, 3}
        \end{pmatrix}^{\!\top}  = \Delta \begin{pmatrix}
            1 & 1 & 0 \\0 & 0 & 1
        \end{pmatrix}^{\!\!\top} \begin{pmatrix} z_{\Delta, 1} & z_{\Delta, 2} \end{pmatrix}^\top, \quad \Delta \in \mathbf{\Delta}  = \left\{ \Delta = \left.\operatorname{diag}(\delta_1, \delta_2,\delta_2) \ \right| \ \|\Delta\|_{2,2} <1, \delta_i \in \Real \right\}.
        % \begin{bmatrix}
        %     w_{\Delta, 1} \\ w_{\Delta, 2} \\ w_{\Delta, 3}
        % \end{bmatrix}  = \Delta \begin{bmatrix}
        %     1 & 0 \\ 1 & 0 \\ 0 & 1
        % \end{bmatrix} \begin{bmatrix} z_{\Delta, 1} \\ z_{\Delta, 2} \end{bmatrix}, \quad \Delta \in \mathbf{\Delta}  = \left\{ \Delta = \left.\operatorname{diag}(\delta_1, \delta_2,\delta_2) \ \right| \ \|\Delta\|_{2,2} <1, \delta_i \in \Real \right\}.
    \end{equation}
\end{subequations}
 which is minimal in terms of the number of uncertain parameters and their repetitions.

\begin{figure}[t]
    \centering
    \includegraphics[width=0.4\columnwidth]{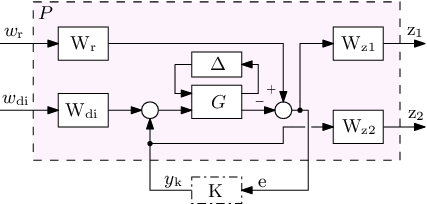}
    \caption{Structure of the generalized plant for controller synthesis in the unbalanced disk example.}
    \label{fig:genUnbalDisk}
\end{figure}
Now we proceed with the construction of the uncertain generalized plant with~\eqref{eq:LFR_unbalanced_disk} to specify the desired closed-loop behaviour of the system. The generalized plant, denoted as $P$ and depicted in Fig.~\ref{fig:genUnbalDisk},
% Now that the unbalanced disk model is approximated with~\eqref{eq:LFR_unbalanced_disk}, we proceed with the construction of the generalized plant to specify the desired closed-loop behaviour of the system. The generalized plant, denoted as $P_{\Sigma}$,
%
%
contains the measured output $u_\mathrm{k}=e$ with $e$ being the tracking error and the controlled input channels $y_\mathrm{k}$ for to-be-synthesized robust controller $K$, the generalized disturbance channels $r$ and $d_i$, and the generalized performance channels $z_1$ and $z_2$. Specifically, $r$ is the reference and $d_i$ is an input disturbance. The weighting filters are chosen as
\begin{equation}\label{eq:unbalDiskfilters}
    W_{\mathrm{z1}}(s) = \frac{0.5012s + 8.3818}{s + 0.8382}, \quad W_{\mathrm{z2}}(s) = \frac{10s + 34.8219}{s + 1101.2}, \quad W_{\mathrm{r}}(s) = \frac{2.282}{s + 0.7216}, \quad W_{\mathrm{di}}(s) = \frac{0.0144}{s + 0.1443},
\end{equation}
where $W_{\mathrm{z1}}$ has low-pass characteristics to ensure good tracking performance at low frequencies, $W_{\mathrm{z2}}$ has high-pass characteristics to enforce roll-off at high frequencies, $W_{\mathrm{r}}$ is a low-pass filter based on the expected bandwidth of the reference signal, and $W_{\mathrm{di}}$ is a low-pass filter that models the expected input disturbances. Then, the generalized plant becomes
\begin{equation}\label{eq:genPlantUnbalDisk}
    \begin{pmatrix}
        z_1 \\ z_2 \\ e
    \end{pmatrix} = P \begin{pmatrix}
        w_{\mathrm{r}} \\ w_{\mathrm{di}} \\ y_\mathrm{k}
    \end{pmatrix}, \quad \text{with} \quad
    P = \begin{pmatrix}
        W_{\mathrm{z1}} W_{\mathrm{r}} & -W_{\mathrm{z1}} G W_{\mathrm{di}} & -W_{\mathrm{z1} }G \\
        0                              & 0                                  & W_{\mathrm{z2}}    \\
        W_{\mathrm{r}}                 & -G W_{\mathrm{di}}                 & -G
    \end{pmatrix}.
\end{equation}
%
%where $\Sigma = \Delta \star G$, $e$ is the input to the controller and $y_k$ is the output of the controller, as illustrated in Fig.~\ref{fig:genUnbalDisk}. 

The grid-based controller synthesis problem is formulated with the controller structure defined by $A_k \in \Real^{3 \times 3}$, $B_k \in \Real^{3 \times 1}$, $C_k \in \Real^{1 \times 3}$, and $D_k = 0$. As indicated in Subsection~\ref{sec:Robustapplication}, the synthesis is executed with \textsc{systune}, and the cost function is formalized as $J(P_\theta, K) = \|T_{w \to z}(\theta, K) \|_\infty$. For comparison purposes, four distinct selections are constructed to approach the synthesis problem. These selections and their corresponding synthesized controllers are obtained as follows:
\begin{enumerate}
    \item \textit{Initial selection $\Theta_0$}: the set of initial grid points is taken as  $\Theta_0 = \{\theta^{(0)}, \ \theta^{(1)}\}$, where $\theta^{(0)} = \operatorname{diag}(\underline{{\delta}}_1, \underline{{\delta}}_2, \underline{{\delta}}_2)$ and $\theta^{(1)} = \operatorname{diag}(\overline{\delta}_1, \overline{\delta}_2, \overline{\delta}_2)$ correspond to the minimum and maximum values of the uncertain parameters, respectively. The controller $K_0$ is synthesized using $P_{\Theta_0}$.
    \item \textit{Dense selection $\Theta_{\mathrm{dense}}$}: each uncertain parameter is sampled at five equidistant points between its minimum and maximum. The controller $K_{\mathrm{dense}}$ is synthesized using $P_{\Theta_{\mathrm{dense}}}$.
          \begin{figure}[b]
              \centering
              \includegraphics[width=1\columnwidth]{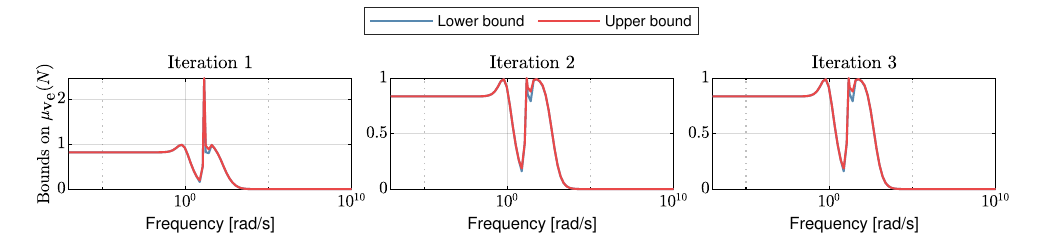}
              \caption{Iterations of the $\mu$-analysis-based grid allocation procedure. Each figure shows the computed bounds of $\mu_{\mathbf{V}_\mathrm{e}}(N)$, which are used to obtain the most informative uncertainty realization w.r.t.\ the closed-loop performance objectives.}
              \label{fig:muAllocation}
          \end{figure}
    \item \textit{$\mu$-analysis-based selection $\Theta_{\mu}$}: the grid points are allocated iteratively using worst-case $\mu$-analysis as outlined in Section~\ref{sec:intro}. The process is initialized with $\Theta_0$ and $K_{0}$. Robust stability is first assessed using the \emph{structured singular value}\cite{doi:10.1049/ip-d.1982.0053,packardComplexStructuredSingular1993} (SSV), denoted by $\mu$, and the small gain theorem, see~\cite{zhouRobustOptimalControl1996,scherer2022robust}.
    If robust stability is not guaranteed, the worst-case point $\theta^\ast$ is added to the grid $\Theta_{\mu} \leftarrow \Theta_\mu \cup \{\theta^\ast \}$. Otherwise, robust performance is evaluated by computing $\mu_{\mathbf{V}_\mathrm{e}}(N(i\omega))$ for all $\omega \in [0, \ \infty]$, where $N(i\omega) = P(i\omega) \star K(i \omega)$ and $\mathbf{V}_\mathrm{e}$ is the extended value set for the considered uncertainty structure:
          \begin{equation}
              \mathbf{V}_\mathrm{e} \coloneq \left\{\operatorname{diag}({V}, \hat{V}) \ | \ V \in \mathbf{V}_{\mathrm{c}}, \ \hat{V} \in \mathbb{C}^{2 \times 2}\right\}.
          \end{equation}
          The point $\theta^\ast$ that maximizes $\mu_{\mathbf{V}_\mathrm{e}}(N(i\omega))$ is then added to $\Theta_\mu$. Next, a new controller is synthesized using $P_{\Theta_\mu}$.
          These steps are repeated iteratively until the desired number grid points is allocated. In this example, three points are allocated with this process.
          The total computation time amounts to 13.27~s, comprising 6.41~s for the execution of the worst-case $\mu$-analysis and 6.86~s for the controller synthesis. Although robust stability is ensured at each iteration, robust performance is only achieved after the first point is allocated, as shown in Fig.~\ref{fig:muAllocation}.
          \begin{figure}[b]
              \centering
              \includegraphics[width=1\columnwidth]{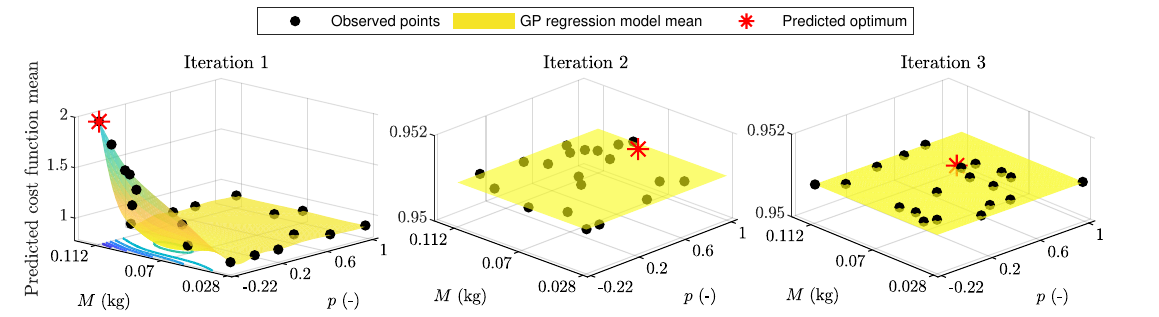}
              \caption{Iterations of the proposed grid point allocation procedure. Each figure shows the predicted mean and optimum of~\eqref{eq:mostInformativeOptim}, which is used to obtain the most informative uncertainty realization w.r.t.\ the closed-loop performance objectives.}
              \label{fig:boIterUnbalDisk}
          \end{figure}
          Then, the controller $K_\mu$ is synthesized using $P_{\Theta_\mu}$.
    \item \textit{Bayesian optimization-based selection $\Theta_{\mathrm{BO}}$}: The fourth set is obtained using the proposed method, where the generalized plant $P$ in~\eqref{eq:genPlantUnbalDisk} and the cost $J$ in~\eqref{eq:RPcost} are used in Algorithms~\ref{generalAlgorithm} and~\eqref{alg:bayesOpt}. Algorithm~\eqref{generalAlgorithm} is initialized with $\Theta_0$ and $K_0$, and the number of iterations is set to  $3$, i.e., $N_\theta =3+2$ grid points are selected as $|\Theta_0|=2$. Algorithm~\eqref{alg:bayesOpt} is initialized with a zero prior mean $m(\theta)=0$, the $C_{5/2}$ kernel~\eqref{kernelc52} is used, and the exploration ratio is set to $\epsilon = 0.3$ with a maximum of $N_{\mathrm{max}} = 20$ iterations. The observation set $\mathbb{D}_N$ is initialized with $N=5$ random points contained in the admissible space. The total computation time of this procedure amounts to 10.19~s, comprising 2.34~s for the execution of Algorithm~~\eqref{alg:bayesOpt} and 7.85~s for the controller synthesis.
    In the first iteration of this procedure, the uncertainty realization allocated by the proposed algorithm coincides with the worst-case realization previously identified by the $\mu$-analysis-based allocation procedure. The predicted mean of the cost function obtained by the GP regression model, together with the optimum selected at each iteration, is illustrated in Fig.~\ref{fig:boIterUnbalDisk}. The controller $K_{\mathrm{BO}}$ is then synthesized using $P_{\Theta_{\mathrm{BO}}}$.
\end{enumerate}
\begin{figure}[b]
    \centering
    \includegraphics[width=0.98\columnwidth]{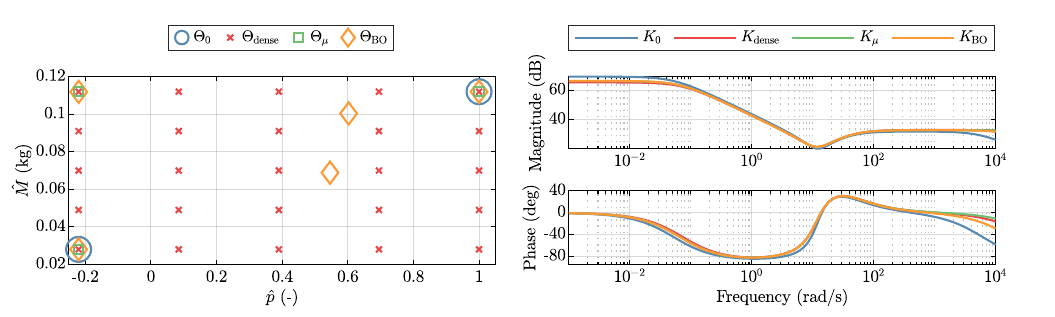}\vspace{-5pt}
    \caption{The left plot shows the points allocated in $\Theta_0$, $\Theta_{\mathrm{dense}}$, $\Theta_{\mu}$, and $\Theta_{\mathrm{BO}}$. The right plot shows the frequency response of $K_0$, $K_{\mathrm{dense}}$, $K_{\mu}$, and $K_{\mathrm{BO}}$, which are synthesized with the sets $P_{\Theta_0}$, $P_{\Theta_{\mathrm{dense}}}$, $P_{\Theta_\mu}$, and $P_{\Theta_{\mathrm{BO}}}$, respectively.}
    \label{fig:gridKcompareUnbalDisk}
\end{figure}

In both $\Theta_\mu$ and $\Theta_{\mathrm{BO}}$, the first point allocated is $\theta = \operatorname{diag}(1, -1, -1)$, identified as the most informative uncertainty point with respect to the closed-loop performance specifications. Once this point is included, additional points yield minor improvements, indicating that the first point provides sufficient information to synthesize a robust structured controller that satisfies the global performance objectives. The grid point selection is visualized in the left panel of Fig.~\ref{fig:gridKcompareUnbalDisk}, where $\theta = \operatorname{diag}(1, -1, -1)$ corresponds to the point in the upper-left corner. This point also appears in $\Theta_{\mathrm{dense}}$, though only due to the nature of the uniform sampling. The frequency responses in the right plot of Fig.~\ref{fig:gridKcompareUnbalDisk} reveal that $K_{\mathrm{dense}}$, $K_\mu$, and $K_{\mathrm{BO}}$ behave nearly identically, despite being synthesized from different sets of local models, suggesting that the contribution of most points in $\Theta_{\mathrm{dense}}$ is negligible.
\begin{figure}[b]
    \centering
    \includegraphics[width=1\columnwidth]{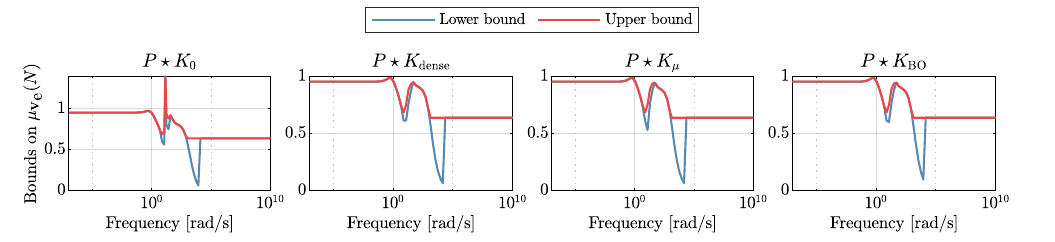}
    \caption{\emph{A posteriori} robust performance analysis of the generalized plant $P$ in feedback with $K_0$, $K_{\mathrm{dense}}$, $K_\mu$, and $K_{\mathrm{BO}}$ for the unbalanced disc system.}
    \label{fig:postMuUnbalDisk}
\end{figure}

\begin{figure}[b]
    \centering
    \includegraphics[width=1\columnwidth]{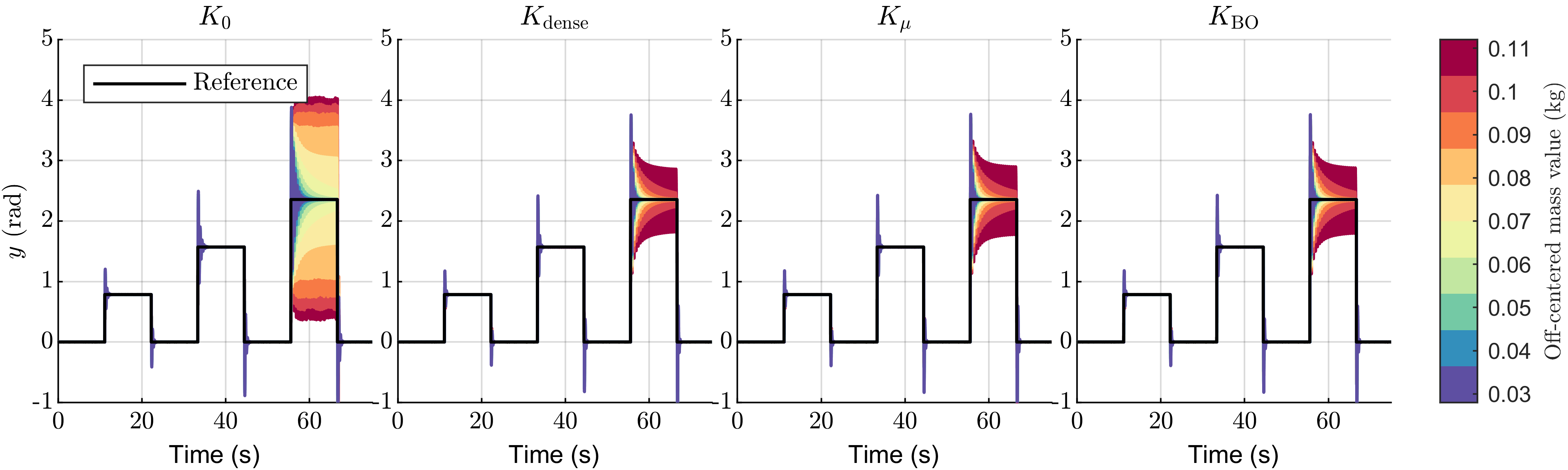}
    \caption{Time-domain simulation results of the nonlinear unbalanced disk system~\eqref{eq:unbalDiskNL} with the $K_0$, $K_{\mathrm{dense}}$, $K_\mu$, and $K_{\mathrm{BO}}$ in feedback configuration. The off-centered mass $M$ is set at different value within the defined uncertainty range at each simulation.}
    \label{fig:NLsimUnbalDisk}
\end{figure}

The synthesized controllers are validated with a posteriori worst-case analysis and nonlinear simulation study. For the worst-case analysis, the bounds on $\mu_{\mathbf{V}_\mathrm{e}}(N(i\omega))$ are computed for each controller. As shown in Fig.~\ref{fig:postMuUnbalDisk}, only $K_{\mathrm{dense}}$, $K_\mu$, and $K_{\mathrm{BO}}$ yield upper bounds whose supremum is below one, thus certifying global robust stability and performance.
The simulation study involves continuous-time simulations of the nonlinear system in~\eqref{eq:unbalDiskNL} with each controller implemented in feedback configuration. The value of the off-centered mass $M$ is varied within its uncertainty range for each simulation, and a challenging reference signal is applied. Additionally, an input disturbance - generated as a discrete-time white-noise with a power spectral density of 0.7 and a zero-order-hold with sampling time of 0.01 seconds, filtered through $W_{\mathrm{di}}$ in~\eqref{eq:unbalDiskfilters} - is injected into the plant. Simulations are run with the variable-step solver \textsc{ode45} of \textsc{Matlab} with default parameters and step-size. The system is initialized with the mass at the downward position with zero velocity, i.e., $x(0) = \operatorname{col}(0, 0)$ and all controller states set to zero. The simulation results, shown in Fig.~\ref{fig:NLsimUnbalDisk}, display that $K_{\mathrm{dense}}$, $K_\mu$, and $K_{\mathrm{BO}}$ offer comparable closed-loop performance across all tested values of ${\bf \Delta}$.
Finally, note that although $K_{\mathrm{dense}}$ exhibits comparable performance to $K_\mu$ and $K_{\mathrm{BO}}$, $K_{\mathrm{dense}}$ is synthesized using 25 local model realizations, whereas only 5 local models are used for $K_\mu$ and $K_{\mathrm{BO}}$. Moreover, the $\mu$-analysis-based allocation procedure used to obtain $\Theta_\mu$ is theoretically guaranteed to identify the worst-case points in the uncertainty space. In contrast, the proposed method offers no such guarantees, as the optimization of~\eqref{eq:mostInformativeOptim} can fail in finding the global optimum. Consequently, the performance achieved with $\Theta_\mu$ can be interpreted as an upper bound for the proposed method. However, the computational burden of the $\mu$-based procedure is significantly higher and often prohibitive in practise. In this sense, the proposed method is computationally efficient and scalable to high-dimensional parameter spaces, as demonstrated in the next example.

\subsection{Satellite with parametric uncertainties and flexible rotating solar arrays\label{sec:satelliteExample}}
Having demonstrated the proposed method on a simple system, we now address a problem of practical relevance in the aerospace industry. Specifically, we consider the multi-objective pointing control problem for a satellite with flexible, rotating solar arrays, subject to several parametric uncertainties, as described in\cite{sanfedinoSatelliteDynamicsToolbox2023}. The satellite dynamics are modeled using an LFR, where the rotation of the solar arrays is treated as a parametric uncertainty. This results in an uncertainty description of the form
\begin{subequations}
    \begin{equation}\label{eq:satelliteUncertainty}
        \Delta \in \mathbf{\Delta} \coloneq \left\{\Delta = \operatorname{diag}(\mathbf{\Delta}_{\mathcal{A}_1},\mathbf{\Delta}_{\mathcal{A}_2},\mathbf{\Delta}_{\mathcal{B}}, \mathbf{\Delta}_{\sigma_4},\mathbf{\Delta}_{\sigma_4},\mathbf{\Delta}_{\sigma_4},\mathbf{\Delta}_{\sigma_4}) \ | \ \|\Delta\|_{2,2}<1  \right\},
        % \Delta \in \mathbf{\Delta} \coloneq \left\{\Delta = \operatorname{diag}(\delta_m I_3, \delta_{xx}, \delta_{xx}, \delta_{xx}, \delta_{\omega_1} I_4, \delta_{\omega_1}, \delta_{\omega_1},)  \right\}.
    \end{equation}
    with
    \begin{equation}
        \begin{aligned}
            \mathbf{\Delta}_{\mathcal{A}_{i}} & \coloneq \left\{\operatorname{diag}(\delta_{\omega_1} I_4, \delta_{\omega_2} I_4, \delta_{\omega_3} I_4)  \right\}, & \mathbf{\Delta}_{\mathcal{B}} & \coloneq \left\{\operatorname{diag}(\delta_m I_3, \delta_{xx}, \delta_{yy}, \delta_{zz})\right\}, & \mathbf{\Delta}_{\sigma_4} & \coloneq \left\{(\delta_{\sigma_4} I_8)\right\}, & \delta_{(\bullet )} & \in \Real,
        \end{aligned}
    \end{equation}
\end{subequations}
where $\delta_{\omega_i}$ represents the uncertainty of the frequency of the $i$-th flexible mode of the solar arrays; $\delta_m$ represents the uncertainty of the mass of the main body of the satellite; $\delta_{xx}$, $\delta_{yy}$, and $\delta_{zz}$ account for the uncertainty of the principal moments of inertia; and $\delta_{\sigma_4}$ expresses the variations of the rotation angle of the solar arrays.
The satellite is actuated by a reaction wheel assembly, whose dynamics are approximated by a second-order low-pass filter. The onboard sensor system includes a star tracker and a gyroscope, modelled as first-order low-pass filters with cut-off frequencies of 8 Hz and 200 Hz, respectively. A 10 ms control-loop delay is approximated by a second-order Padé filter. Additionally, low-frequency orbital disturbances and sensor noise are taken into account.
The to-be-synthesized $3 \times 6$ attitude controller $K(s)$ is structured as a diagonal controller, consisting of a proportional-derivative controller with a first order low-pass filter for each axis. For a complete description of the satellite model, avionics, considered disturbances, controller architecture, and initialization of controller parameters, the reader is referred to \cite{sanfedinoSatelliteDynamicsToolbox2023}. These details are omitted here for the sake of compactness.
\begin{figure}[b]
    \centering
    \includegraphics[width=0.8\columnwidth]{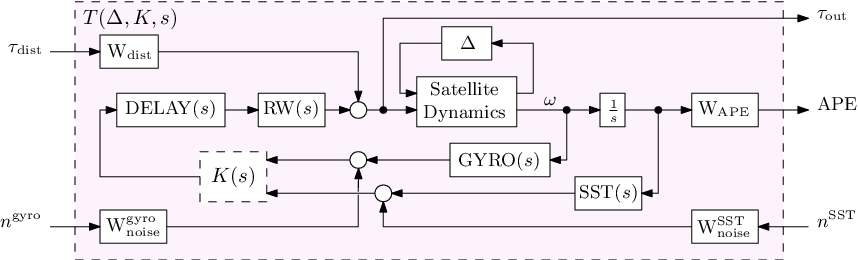}
    \caption{Uncertain closed-loop generalized plant representing the pointing control problem of the satellite with flexible rotating solar arrays.}
    \label{fig:genplantSatellite}
\end{figure}
The controller must satisfy the following requirements for any $\Delta \in \mathbf{\Delta}$:
% \begin{enumerate}[label=(Req. \arabic*)]
\begin{itemize}
    \item\label{Req1} \textbf{(Req. 1)} The \emph{absolute pointing error} (APE) must be less than $[4 \ 4 \ 20]^\top \frac{\pi}{180} 10^{-3}$ rad.
    \item\label{Req2} \textbf{(Req. 2)} The $\mathcal{H}_\infty$-norm of the input sensitivity function must be upper bounded by $\gamma = 1.5$.
    \item\label{Req3} \textbf{(Req. 3)} Minimize the variance of the torque applied by the reaction wheel on the spacecraft in response to the measurement noise of the star tracker and gyro sensors.
\end{itemize}
% \begin{enumerate}[label=Req. \arabic*:, labelsep=.5em, leftmargin=5em]
%     \item\label{Req1} Test
%     \item\label{Req2} Test 2
%     \item\label{Req3} Test 3
% \end{enumerate}
The uncertain generalized plant $T(\Delta, K, s) = \Delta \star P(s) \star K(s)$ and the closed-loop performance specifications used for the controller synthesis are shown in Fig.~\ref{fig:genplantSatellite}. Next, we combine the three performance requirements in the following cost function:
\begin{equation}\label{eq:satelliteCost}
    J  = \max_{\Delta} \underbrace{\|T_{\tau_{\mathrm{dist}} \to \mathrm{APE}}(\Delta, K, s)\|_\infty}_{J_1} + \underbrace{\|T_{\tau_{\mathrm{dist}} \to \tau_{\mathrm{out}}}(\Delta, K, s)\|_\infty}_{J_2} + \underbrace{\|T_{\nnoise \to \tau_{\mathrm{out}}}(\Delta, K, s)\|_2}_{J_3},
\end{equation}
where $\nnoise = [n^{\mathrm{gyro}} \ n^{\mathrm{SST}}]^\top$. Note that our treatment of the performance requirements differs from that in~\cite{sanfedinoSatelliteDynamicsToolbox2023}. Here, the individual requirements are combined into a single cost function. This reformulation aims to reduce the discontinuities in the cost function, thereby enhancing the effectiveness of the optimization in~\eqref{eq:mostInformativeOptim}.
A second key difference from~\cite{sanfedinoSatelliteDynamicsToolbox2023} is that, rather than directly supplying the uncertain plant to \textsc{systune}, we treat the controller synthesis problem in a gridded fashion:
\begin{equation}\label{eq:satelliteSynthesis}
    K^\ast = \argmin_K \left\{\|T_{\tau_{\mathrm{dist}} \to \mathrm{APE}}(\theta^{(i)}, K, s)\|_\infty + \|T_{\tau_{\mathrm{dist}} \to \tau_{\mathrm{out}}}(\theta^{(i)}, K, s)\|_\infty + \|T_{\nnoise \to \tau_{\mathrm{out}}}(\theta^{(i)}, K, s)\|_2\right\}_{i=1}^{N_\theta}, \quad \theta^{(i)} \in \Delta,
\end{equation}
where we consider the minimization of the sum of the three performance objectives for convenience.

With the cost function~\eqref{eq:satelliteCost} and the synthesis problem~\eqref{eq:satelliteSynthesis} defined, a selection is allocated using the proposed method. Algorithm~\ref{generalAlgorithm} is initialized with the grid $\Theta_0 = \{ \theta^{(1)}, \theta^{(2)}\}$, where $\theta^{(1)} = \operatorname{diag}(-1,\ldots,-1)$ and $\theta^{(2)} = \operatorname{diag}(1,\ldots,1)$,
% \begin{equation*}
%  \Theta_0 = \{ \theta^{(1)}, \theta^{(2)}\}, \qquad  \theta^{(1)} = \operatorname{diag}(-1,\ldots,-1), \qquad
% \theta^{(2)} = \operatorname{diag}(1,\ldots,1),
% \end{equation*}
the controller $K_0$ synthesized using $P_{\Theta_0}$ and $N_\theta = 8$.
For Algorithm~\ref{alg:bayesOpt}, the prior mean of the GP model is chosen as $m(\theta)=0$, and the Matérn-$5/2$ kernel~\eqref{kernelc52} is employed. The exploration ratio of the acquisition function is set to $\epsilon = 0.5$, the maximum number of BO iterations is fixed to $N_{\mathrm{max}} = 30$, and the observation set $\mathbb{D}_N$ is initialized with $N=20$ random points within the admissible space.
The resulting set $\Theta_{\mathrm{BO}}$ with $|\Theta_{\mathrm{BO}}|=8$ is then used to synthesize $K_{\mathrm{BO}}$ with $P_{\Theta_{\mathrm{BO}}}$.
The total computation time amounts to 488.3~s, comprising 121.7~s for optimizing~\eqref{eq:mostInformativeOptim} via Algorithm~\ref{alg:bayesOpt} and 366.6~s for controller synthesis using \textsc{Systune}\footnote{In this case, the \emph{randomstart} option of \textsc{systune} is set to 5.}.

% With the cost function~\eqref{eq:satelliteCost} and synthesis problem~\eqref{eq:satelliteSynthesis} defined, we allocate a grid using the proposed method. Algorithm~\ref{generalAlgorithm} is initialized with an initial grid $\Theta_0$, constructed by selecting the two boundary points of the uncertainty set $\theta^{(1)} = \operatorname{diag}(-1, -1, -1, -1, -1, -1, -1, -1)$ and $\theta^{(2)} = \operatorname{diag}(1, 1, 1, 1, 1, 1, 1, 1)$, and the initial controller $K_0$ synthesized using $\Theta_0$. The number of points to allocate is set to $N_\theta = 8$.
% For Algorithm~\ref{alg:bayesOpt}, the prior mean is set to $m(\theta)=0$, the $C_{5/2}$ kernel~\eqref{kernelc52} is employed, an exploration ratio of $\epsilon = 0.5$ is set for the acquisition function, a maximum number of $N_{\mathrm{max}} = 30$ BO iterations is used, and the observation set $\mathbb{D}_N$ is initialized with $N=20$ random points within the admissible space.
% Under these settings, $N_\theta = 8$ allocated points are appended to $\Theta_0$, resulting in the set $\Theta_{\mathrm{BO}}$. The total computation time is 488.3 s, comprising 121.7 s for optimizing~\eqref{eq:mostInformativeOptim} via Algorithm~\ref{alg:bayesOpt}, and 366.6 s for controller synthesis using \textsc{Systune}\footnote{In this case, the \emph{randomstart} option of \textsc{systune} is set to 5.}. Then, the final controller $K_{\mathrm{BO}}$ is synthesized using $P_{\mathrm{BO}}$.
%
\begin{figure}[b]
    \centering
    \includegraphics[width=1\columnwidth]{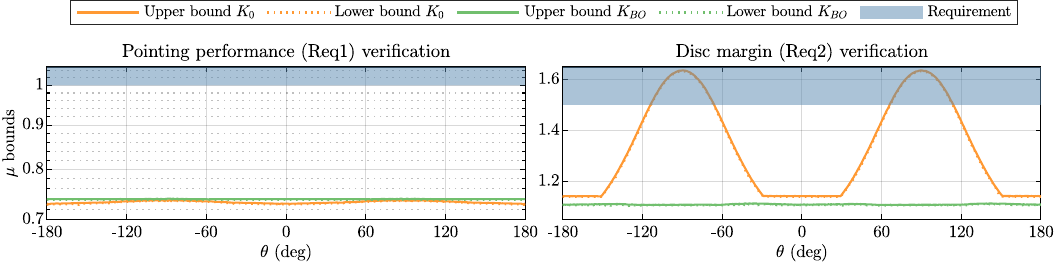}
    \caption{Posterior worst-case analysis with the controllers $K_0$ and $K_{\mathrm{BO}}$ for the satellite control problem. The $\mu$-bounds for a grid of  361 equidistant values of $\delta_{\sigma_4}$ for the pointing and for the disc margin requirements are shown in the left and right figures, respectively.}
    \label{fig:postMuSatellite}
\end{figure}
A posteriori worst-case analysis is performed to verify whether the controllers $K_0$ and $K_{\mathrm{BO}}$ satisfy the requirements (Req. 1) and (Req. 2). This analysis employs the \textsc{Matlab} built-in function \textsc{wcgain}\footnote{We use the \textsc{wcgain} option \emph{`a'}, which computes the upper bound on the structured singular values to the greatest accuracy by means of an LMI procedure.}.
As discussed in~\cite{sanfedinoSatelliteDynamicsToolbox2023}, conventional worst-case analysis tools fail in the computation of the bounds on the structured singular value due to the 32 repetitions of $\delta_{\sigma_4}$ contained in $\mathbf{\Delta}$. To circumvent this issue, the worst-case analysis must be conducted by sampling $\delta_{\sigma_4}$ across a range of values. To this end, we sample $\delta_{\sigma_4}$ over a fine grid of 361 points, corresponding to an angular position of the solar arrays from $-180^\circ$ to $180^\circ$ in 1-degree increments.
The computation times required to test (Req. 1) and (Req. 2) are $2576$ s and $14517$ s, respectively. As shown in Fig.~\ref{fig:postMuSatellite}, $K_{\mathrm{BO}}$ satisfies both requirements across all sampled values of $\delta_{\sigma_4}$, while $K_0$ fails to meet (Req. 2) in the intervals $(-110^\circ, -70^\circ)$ and $(70^\circ, 110^\circ)$.

The $\mu$-analysis-based allocation method detailed in Subsection~\ref{sec:unbalanced_disk} is not applied for this case. As previously discussed, the number of repetitions of $\delta_{\sigma_4}$ in~\eqref{eq:satelliteUncertainty} render the full $\mathbf{\Delta}$ block intractable for the existing worst-case tools (they can only be employed in a gridded fashion). Consequently, the guarantees on finding the true worst-case point are lost, which may lie between grid samples.
The a posteriori analysis, presented in Fig.~\ref{fig:postMuSatellite}, required a total runtime of $17093$ s to validate both requirements. Applying the $\mu$-analysis-based allocation method would demand a comparable computational effort per grid point, leading to an estimated runtime of approximately $28.5$ h to allocate 6 points. In contrast, the proposed method completes the entire allocation in just $121.7$ s.

Lastly, to the authors' knowledge, there is no existing tool that can be used to provide analytical guarantees for the variance requirement (Req. 3).
% Therefore, worst-case analysis tools can not be used to find the most informative point or to validate this criterion. 
In~\cite{sanfedinoSatelliteDynamicsToolbox2023}, its minimization is left to the heuristic worst-case parametric and geometric configurations automatically detected by \textsc{Systune}. In contrast, the proposed method explicitly incorporates (Req. 3) into the cost function~\eqref{eq:satelliteCost}, enabling the allocation of grid points that are also informative w.r.t.\ the variance requirement.

\subsection{Two-degrees of freedom robotic arm\label{sec:robotArmExample}}
In this last example, we test the developed method for the LPV controller design problem of a two-degrees of freedom planar robotic arm detailed in~\cite{kwiatkowskiLPVControl2DOF2005}. The equations of motion of the robotic arm can be described in the nonlinear state-space form
\begin{subequations}\label{eq:robotArmNL}
    \begin{equation}
        \Sigma \coloneq \left\{ \begin{aligned}
            \dot{x}(t) & = \begin{pmatrix}
                               \dot{q}(t) \\
                               M^{-1}(q(t)) \left(n \, u(t) - C(q(t), \dot{q}(t)) \right) + g(q(t))
                           \end{pmatrix}, \\
            y(t)       & = \begin{pmatrix}
                               I_2 & 0
                           \end{pmatrix}x(t),
        \end{aligned}\right.
    \end{equation}
    with
    \begin{align*}
        M(q(t)) & = \begin{pmatrix}
                  a                  & b \cos_{\Delta}(t) \\
                  b \cos_{\Delta}(t) & c
              \end{pmatrix}, & g(q(t)) & = \begin{pmatrix}
                                         -d \sin(q_1(t)) \\ -e \sin(q_2(t))
                                     \end{pmatrix}, & C(q(t), \dot{q}(t)) & = \begin{pmatrix}
                                                                b \sin_{\Delta}(t) \, \dot{q}_2(t)^2 + f \dot{q}_1(t) \\
                                                                -b \sin_{\Delta}(t) \, \dot{q}_1(t)^2 + f\left(\dot{q}_2(t) - \dot{q}_1(t) \right)
                                                            \end{pmatrix},
    \end{align*}
\end{subequations}
where $\cos_{\Delta}(t) = \cos \left(q_1(t) - q_2(t) \right)$, $\sin_{\Delta}(t) = \sin \left(q_1(t) - q_2(t) \right)$, $q(t) = \left(q_1(t) \ q_2(t)\right)^\top$ are the angles, $x(t) = \left(q_1(t) \ q_2(t) \ \dot{q}_1(t)\ \dot{q}_2(t)\right)^\top$ is the state vector and $u(t) = \left(\tau_1(t) \ \tau_2(t)\right)^\top$ are the input motor torques. The values of the physical parameters of the robotic arm are $(a, \ b, \ c, \ d, \ e, \ f, \ n) = (5.6794, \ 1.473, \ 1.7985, \ 0.4, \ 0.4, \ 2, \ 1)$. The LPV model of the robotic arm, based on the LPV embedding in\cite{kwiatkowskiLPVControl2DOF2005}, is given by
\begin{subequations}
    \begin{equation}
    G_{p(t)} \coloneq \left\{\begin{aligned}
            \dot{x}(t) & = A(p(t))x(t) + B(p(t))u(t), \\
            y(t)       & = Cx(t) + Du(t),
        \end{aligned}\right.
    \end{equation}
    where
    \begin{equation}
        \begin{aligned}
            A(p) & = \begin{pmatrix}
                         0 & 0 & 1 & 0 \\ 0 & 0 & 0 & 1 \\ c  d  p_3 & - b  e  p_4 & p_5 & b  p_6 \\ -b  d  p_7 & a  e  p_8 & p_9 & p_{10}
                     \end{pmatrix}; & B(p) & = \begin{pmatrix}
                                                   0 & 0 \\ 0 & 0 \\ c  n  p_1 & -b  n p_2 \\ -b  n  p_2 & a  n  p_1
                                               \end{pmatrix}; & C & = \begin{pmatrix}
                                                                          I_2 & 0
                                                                      \end{pmatrix}; & D & = 0,
        \end{aligned}
    \end{equation}
    and the scheduling map $p = \eta(x)$, with $\eta : \Real^4 \to \Real^{10}$, is given by
    \begin{equation}\label{eq:robotSchedMap}
        \eta(x)=\left(\begin{array}{l}
                1 / h                                                                                     \\
                \cos _{\Delta}(t) / h                                                                     \\
                \operatorname{sinc}\left(x_1\right) / h                                                   \\
                \cos _{\Delta} \operatorname{sinc}\left(x_2\right) / h                                    \\
                \left(-b^2 \sin _{\Delta} \cos _{\Delta} x_3-\left(c+b \cos _{\Delta}\right) f\right) / h \\
                \left(-c \sin _{\Delta} x_4+\cos _{\Delta} f\right) / h                                   \\
                \cos _{\Delta} \operatorname{sinc}\left(x_1\right) / h                                    \\
                \operatorname{sinc}\left(x_2\right) / h                                                   \\
                \left(a b \sin _{\Delta} x_3+f\left(a+b \cos _{\Delta}\right)\right) / h                  \\
                \left(b^2 \sin _{\Delta} \cos_{\Delta} x_4-a f\right) / h
            \end{array}\right); \quad h = a c - b^2(\cos_{\Delta})^2.
    \end{equation}
\end{subequations}

\begin{figure}[b]
    \centering
    \includegraphics[width=0.4\columnwidth]{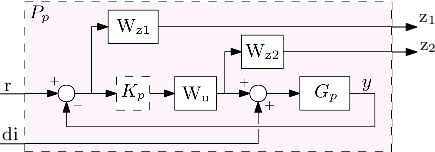}
    \caption{Generalized plant structure for the LPV controller synthesis of the robotic arm.}
    \label{fig:genRobot}
\end{figure}

Next, the closed-loop performance objectives are based in\cite{rizviKernelBasedPCAApproach2016}, where the goal is to achieve reference tracking of the two angles $q_1$ and $q_2$. For this end, based on~\cite{koelewijnSchedulingDimensionReduction2020}, we construct the generalized plant shown in Fig.~\ref{fig:genRobot},
where $r$ is the reference signal to be tracked, $d$ represents an input disturbance, and $[z_1 \ z_2]^\top$ are the generalized performance channels. The weighting filters are
\begin{subequations}
    \begin{equation}
        \begin{aligned}
            W_{\mathrm{z1}}(s) & =\operatorname{diag}(W_1, \, W_1)(s); & W_{\mathrm{z2}} & =\operatorname{diag}(W_2, \, W_2); & W_{\mathrm{u}}(s) & =\operatorname{diag}(W_3, \, W_3)(s),
        \end{aligned}
    \end{equation}
    with
    \begin{equation}
        \begin{aligned}
            W_1(s) & =\frac{0.5s + 5}{s + 5 \cdot 10^{-5}}; & W_2 & = 3 \cdot 10^{-3}; & W_3(s) & = \frac{1\cdot10^3}{s + 1\cdot10^3},
        \end{aligned}
    \end{equation}
\end{subequations}
where $W_1$ includes low-pass characteristics to ensure good tracking performance at low frequencies, $W_2$ is a constant gain to limit the motor torques. 
The weighting filter $W_3$ is included as a strictly proper low-pass filter to restrict high-frequent control action and to reduce its sensitivity noise and high-frequent disturbances.
\begin{figure}[t]
    \centering
    \includegraphics[width=1\columnwidth]{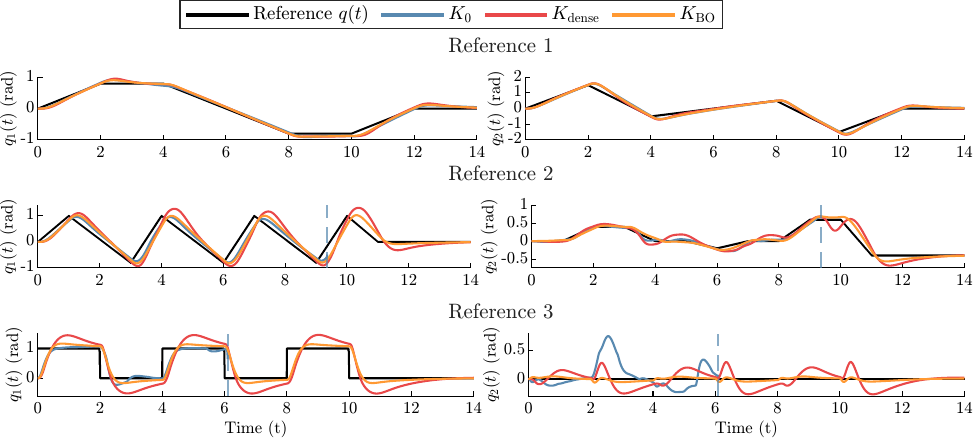}
    \caption{Time-domain simulation results of the nonlinear model of the robotic arm~\eqref{eq:robotArmNL} with the LPV controllers $K_0$, $K_{\mathrm{dense}}$, and $K_{\mathrm{BO}}$ in feedback configuration.}
    \label{fig:robotArmSim}
\end{figure}

We now apply the proposed method to allocate a grid for synthesizing an LPV controller that satisfies these closed-loop performance objectives. The tracking performance is assessed using three reference trajectories, illustrated in Fig.~\ref{fig:robotArmSim}. 
The first trajectory, denoted as \mbox{`Reference 1'}, is used together with~\eqref{eq:robotSchedMap} to determine the admissible scheduling space by generating a trajectory of the scheduling variables. 
Note that the angular rate of this reference is also required to obtain the scheduling trajectories. The bounds $\underline{p}_i$ and $\overline{p}_i$ of the admissible scheduling space are then defined as the minimum and maximum values of this scheduling trajectory. \mbox{`Reference 2'} and \mbox{`Reference 3'} are specifically designed to test the extrapolation capabilities and robustness of the LPV controllers, as they involve angular positions and rates that are mapped outside the admissible scheduling space.
Algorithm~\ref{generalAlgorithm} is initialized with the selection $\Theta_0 = \{\underline{p}_i, \ {(\underline{p}_i + \overline{p}_i)}/{2}, \ \overline{p}_i\}$, the corresponding LPV controller $K_0(p)$ synthesized with $\Theta_0$ and {$N_\theta = 16$}. To initialize Algorithm~\ref{alg:bayesOpt}, the prior mean is set to $m(\theta) = 0$, the $C_{5/2}$ kernel is employed, an exploration ratio of {$\epsilon = 0.7$} is used, the maximum number of BO iterations is fixed to $N_\mathrm{max} = 40$ and the observation set $\mathbb{D}_N$ is obtained by drawing $N=20$ random points within the admissible scheduling space.
Under these settings, $\Theta_{\mathrm{BO}}$ is obtained by appending $N_\theta = 16$ points to $\Theta_0$, and is subsequently used to synthesize the LPV controller $K_{\mathrm{BO}}(p)$.  All LPV controller syntheses are performed using the gridded approach indicated in Subsection~\ref{sec:LPVapplication}. 
For comparison, an additional LPV controller $K_{\mathrm{dense}}(p)$ is synthesized using a dense selection $\Theta_{\mathrm{dense}}$, obtained by sampling the admissible scheduling space with two equidistant points per dimension, resulting in a total of $N_\theta = 2^{10} = 1024$ grid points.

The controllers $K_0(p)$, $K_{\mathrm{dense}}(p)$, and $K_{\mathrm{BO}}(p)$ are then interconnected with the nonlinear robotic arm model~\eqref{eq:robotArmNL}, and their tracking performance is evaluated using the three reference trajectories. The simulations are executed using \textsc{Matlab}'s built-in variable-step solver \textsc{ode15s} with the default parameters and step-size. The states associated with the representation of the system and of the LPV controllers are initialized at zero.
The simulation results, shown in Fig.~\ref{fig:robotArmSim}, display that $K_0(p)$ fails to stabilize the system for `Reference 2' and `Reference 3', and the numerical solver halts due to singularities in the solutions. $K_{\mathrm{dense}}(p)$, despite being synthesized using a significantly larger grid, exhibits a notorious performance degradation for `Reference 2' and `Reference 3'. In contrast, $K_{\mathrm{BO}}(p)$ achieves consistently good tracking performance across all references.
Lastly, the tracking performance is quantitatively assessed by computing the root-mean-square error between the system outputs and the reference trajectories, which are given in Table~\ref{tab:robotArmSim}.
\begin{table}[t]
    \centering
    \caption{Root-mean-square error between the system output trajectories and the reference trajectories.}\label{tab:robotArmSim}
    \renewcommand{\arraystretch}{1.01}
    \begin{tabularx}{0.95\textwidth}{c|YY|YY|YY}
        % \cline{2-7}
                             & \multicolumn{2}{c|}{$\mathbf{K_0}$} & \multicolumn{2}{c|}{$\mathbf{K_{\mathrm{dense}}}$} & \multicolumn{2}{c}{$\mathbf{K_{\mathrm{BO}}}$}                                                       \\
        \cline{2-7}
                             & $\mathbf{q_1}$                      & $\mathbf{q_2}$                                     & $\mathbf{q_1}$                                 & $\mathbf{q_2}$  & $\mathbf{q_1}$  & $\mathbf{q_2}$  \\
        \midrule
        \textbf{Reference 1} & 0.0707                              & 0.2002                                             & 0.0784                                         & \textbf{0.1362} & \textbf{0.0686} & 0.1562          \\
        % \hline
        \textbf{Reference 2} & $\infty$                            & $\infty$                                           & 0.4024                                         & 0.1338          & \textbf{0.3165} & \textbf{0.0791} \\
        % \hline
        \textbf{Reference 3} & $\infty$                            & $\infty$                                           & 0.4608                                         & 0.1339          & \textbf{0.4136} & \textbf{0.0268} \\
        % \hline
    \end{tabularx}
\end{table}

\section{Conclusion}\label{Sec:conclusion}
This paper introduced a novel grid point allocation method for gridded robust and LPV controller design. In contrast to traditional approaches, which rely on engineering insight, dense sampling or worst-case analysis, the proposed method identifies the most informative grid point with respect to the user-specified closed-loop performance objectives in a computationally efficient manner. To this end, Bayesian optimization is employed to optimize a cost function that captures both the performance and robustness objectives of the closed-loop system while minimizing the number of required local model evaluations.
The effectiveness of the proposed method is demonstrated through several case studies, including an unbalanced disk system, a satellite pointing control problem, and the control of a two-degrees of freedom robotic arm. In the robust control setting, the proposed approach outperforms $\mu$-analysis-based allocation methods, specifically in cases where the computation of the bounds on the structured singular value is unreliable due to the complexity of the uncertainty description or the existence of repeated real parametric uncertainties, as demonstrated in the satellite pointing example of Subsection~\ref{sec:satelliteExample}. Moreover, the method enables the explicit inclusion of performance objectives such as variance minimization, for which there exist no tools that provide analytical guarantees, up to the authors' knowledge. In the LPV context, where worst-case analysis tools are not directly applicable, the method is shown to be effective in identifying the most informative scheduling points for LPV controller synthesis, as shown in the robotic arm example in Subsection~\ref{sec:robotArmExample}. In addition, the proposed radial basis function interpolation scheme enables the online implementation of gridded LPV controllers synthesized from scattered evaluations of the local models.

Future research directions include the study of formal \emph{global} stability and performance guarantees for controllers synthesized using the proposed method, as the current certification relies on a posteriori analysis. Furthermore, alternative optimization strategies for the cost function, such as gradient-based techniques, adaptive sampling or reinforcement learning, can be explored to enhance the computational efficiency, scalability and robustness of the proposed allocation procedure.
%%%%%%%%%%%%%%%%%%%%%%%%%%%%%%%%%%%%%%%%%%%%%%%%%%%%%%%%%

%%%%%%%%%%%%%%% BACK MATTER %%%%%%%%%%%%%%%%%%%%%%%%%%%%%
\section*{Acknowledgments}
We thank Dr. Valentin Preda from ESA for the fruitful discussions that contributed to the development of this paper.
%%%%%%%%%%%%%%%%%%%%%%%%%%%%%%%%%%%%%%%%%%%%%%%%%%%%%%%%%

\bibliography{BO-grid-point-allocation}

\providecommand{\noopsort}[1]{}
\begin{thebibliography}{10}

\bibitem{toth2010}
T{\'o}th R. {\it Modeling and identification of linear parameter-varying systems}.
\newblock Germany: Springer; 2010.

\bibitem{https://doi.org/10.1002/rnc.704}
Balas GJ. Linear, Parameter-Varying Control and Its Application to a Turbofan Engine.  {\it International Journal of Robust and Nonlinear Control. }2002;12(9):763--796.

\bibitem{mohammadpourControlLinearParameter2012a}
Mohammadpour J, Scherer CW. {\it Control of {{Linear Parameter Varying Systems}} with {{Applications}}}.
\newblock Boston, US: Springer; 2012.

\bibitem{bachnasReviewDatadrivenLinear2014}
Bachnas AA, T{\'o}th R, Ludlage JHA, Mesbah A. A Review on Data-Driven Linear Parameter-Varying Modeling Approaches: {{A}} High-Purity Distillation Column Case Study.  {\it Journal of Process Control. }2014;24(4):272--285.

\bibitem{hoffmannSurveyLinearParameterVarying2015}
Hoffmann C, Werner H. A {{Survey}} of {{Linear Parameter-Varying Control Applications Validated}} by {{Experiments}} or {{High-Fidelity Simulations}}.  {\it IEEE Transactions on Control Systems Technology. }2015;23(2):416--433.

\bibitem{kwiatkowskiAutomatedGenerationAssessment2006}
Kwiatkowski A, Boll MT, Werner H. Automated {{Generation}} and {{Assessment}} of {{Affine LPV Models}}.  In: Proceedings of the 45th {{IEEE Conference}} on {{Decision}} and {{Control}}:6690--6695; 2006; San Diego, US.

\bibitem{decaignyInterpolationBasedModelingMIMO2011a}
Caigny J, Camino JF, Swevers J. Interpolation-{{Based Modeling}} of {{MIMO LPV Systems}}.  {\it IEEE Transactions on Control Systems Technology. }2011;19(1):46--63.

\bibitem{sadeghzadehAffineLinearParametervarying2020}
Sadeghzadeh A, Sharif B, T{\'o}th R. Affine Linear Parameter-varying Embedding of Non-linear Models with Improved Accuracy and Minimal Overbounding.  {\it IET Control Theory \& Applications. }2020;14(20):3363--3373.

\bibitem{koelewijnLearningReducedOrderLinear2023a}
Koelewijn PJW, Sing R, Seiler P, T{\'o}th R. Learning {{Reduced-Order Linear Parameter-Varying Models}} of {{Nonlinear Systems}}.  {\it IFAC-PapersOnLine. }2024;58(15):265--270.

\bibitem{oluchaAutomatedLinearParameterVarying2025}
Olucha EJ, Koelewijn PJW, Das A, T{\'o}th R. Automated {{Linear Parameter-Varying Modeling}} of {{Nonlinear Systems}}: {{A Global Embedding Approach}}.  {\it IFAC-PapersOnLine. }2025;59(15):49-54.

\bibitem{oluchaReductionLinearParameterVarying2024b}
Olucha EJ, Terzin B, Das A, Tóth R. {\it On the reduction of Linear Parameter-Varying State-Space models. } arXiv:2404.01871 [Preprint]; 2024.

\bibitem{9568982}
Apkarian P, Cabinet P. Erratum to "A Convex Characterization of Gain-Scheduled {{H}}{\textsubscript{{$\infty$}}} Controllers".  {\it IEEE Transactions on Automatic Control. }1995;40(9):1681--1681.

\bibitem{HJARTARSON2015139}
Hjartarson A, Seiler P, Packard A. {{LPVTools}}: A Toolbox for Modeling, Analysis, and Synthesis of Parameter Varying Control Systems.  {\it IFAC-PapersOnLine. }2015;48(26):139--145.

\bibitem{zhouRobustOptimalControl1996}
Zhou K, Doyle JC, Glover K. {\it Robust and Optimal Control}.
\newblock US: Prentice Hall; 1996.

\bibitem{scherer2022robust}
Scherer C. {\it Theory of Robust Control. } Lecture notes; 2022.

\bibitem{4789992}
Doyle J, Glover K, Khargonekar P, Francis B. State-Space Solutions to Standard {{H}}{\textsubscript{2}} and {{H}}{\textsubscript{{$\infty$}}} Control Problems.  In: 1988 American Control Conference:1691--1696; 1988.

\bibitem{POSTLETHWAITE200727}
Postlethwaite I, Turner MC, Herrmann G. Robust Control Applications.  {\it Annual Reviews in Control. }2007;31(1):27--39.

\bibitem{liu2016robust}
Liu KZ, Yao Y. {\it Robust Control: Theory and Applications}.
\newblock Wiley Publishing; 2016.

\bibitem{5612823}
Roos C. A Practical Approach to Worst-Case {{H}}{\textsubscript{{$\infty$}}} Performance Computation.  In: 2010 {{IEEE}} International Symposium on Computer-Aided Control System Design:380--385; 2010.

\bibitem{packardComplexStructuredSingular1993}
Packard A, Doyle J. The Complex Structured Singular Value.  {\it Automatica. }1993;29(1):71--109.

\bibitem{theodoulisGainScheduledAutopilot2008}
Theodoulis S, Duc G. Gain {{Scheduled Autopilot Synthesis}} for an {{Atmosphere Re-Entry Vehicle}}.  In: {{AIAA Guidance}}, {{Navigation}} and {{Control Conference}} and {{Exhibit}}; 2008; Honolulu, US.

\bibitem{fleischmannSystematicLPVLFR2016a}
Fleischmann S, Theodoulis S, Laroche E, Wallner E, Harcaut JP. A {{Systematic LPV}}/{{LFR Modelling Approach Optimized}} for {{Linearised Gain Scheduling Control Synthesis}}.  In: {{AIAA Modeling}} and {{Simulation Technologies Conference}}; 2016; San Diego, US.

\bibitem{sanfedinoExperimentalValidationHigh2019a}
Sanfedino F. Experimental Validation of a High Accuracy Pointing System.
\newblock PhD thesisUniversity of Toulouse / ISAE-SUPAERO2019.

\bibitem{shahriariTakingHumanOut2016a}
Shahriari B, Swersky K, Wang Z, Adams RP, Freitas N. Taking the Human out of the Loop: A Review of Bayesian Optimization.  {\it Proceedings of the IEEE. }2016;104(1):148--175.

\bibitem{Frazier:2018iqo}
Frazier PI. {\it A {{Tutorial}} on {{Bayesian Optimization}}. } arXiv.1807.02811 [Preprint]; 2018.

\bibitem{brochu2010}
Brochu E, Cora VM, Freitas N. {\it A {{Tutorial}} on {{Bayesian Optimization}} of {{Expensive Cost Functions}}, with {{Application}} to {{Active User Modeling}} and {{Hierarchical Reinforcement Learning}}. } arXiv:1012.2599 [Preprint]; 2010.

\bibitem{brosigDataefficientAutotuningBayesian2020}
{Neumann-Brosig} M, Marco A, Schwarzmann D, Trimpe S. Data-Efficient {{Auto-tuning}} with {{Bayesian Optimization}}: {{An Industrial Control Study}}.  {\it IEEE Transactions on Control Systems Technology. }2020;28(3):730--740.

\bibitem{Zhou95}
Zhou K., Doyle J.~C., Glover K.. {\it Robust and Optimal Control}.
\newblock Prentice-Hall; 1995.

\bibitem{briat2015}
Briat C. {\it Linear Parameter-Varying and Time-Delay Systems: Analysis, Observation, Filtering \& Control}.
\newblock Advances in Delays and DynamicsSpringer Berlin, Heidelberg; 2015.

\bibitem{hoffmannSurveyLinearParameterVarying2015a}
Hoffmann C, Werner H. A {{Survey}} of {{Linear Parameter-Varying Control Applications Validated}} by {{Experiments}} or {{High-Fidelity Simulations}}.  {\it IEEE Transactions on Control Systems Technology. }2015;23(2):416--433.

\bibitem{rasmussenGaussianProcessesMachine2006a}
Rasmussen CE. {\it Gaussian Processes for Machine Learning}.
\newblock Adaptive Computation and Machine LearningCambridge: MIT Press; 2006.

\bibitem{bishopPatternRecognitionMachine2016}
Bishop CM. {\it Pattern {{Recognition}} and {{Machine Learning}}}.
\newblock New York, US: Springer; 2016.

\bibitem{wilson2018maximizing}
Wilson J, Hutter F, Deisenroth M. Maximizing Acquisition Functions for {{Bayesian}} Optimization.  {\it Advances in neural information processing systems. }2018;31.

\bibitem{koelewijnAnalysisControlNonlinear2023}
Koelewijn PJW. Analysis and Control of Nonlinear Systems with Stability and Performance Guarantees: {{A}} Linear Parameter-Varying Approach.
\newblock PhD thesisEindhoven University of Technology2023.

\bibitem{DENBOEF2021385}
Boef P, Cox PB, T{\'o}th R. {{LPVcore}}: {{MATLAB}} Toolbox for {{LPV}} Modelling, Identification and Control.  {\it IFAC-PapersOnLine. }2021;54(7):385--390.

\bibitem{amidrorScatteredDataInterpolation2002}
Amidror I. Scattered Data Interpolation Methods for Electronic Imaging Systems: A Survey.  {\it Journal of Electronic Imaging. }2002;11:76 -- 157.

\bibitem{apkarianNonsmoothSynthesis2006}
Apkarian P, Noll D. Nonsmooth {{H}}{\textsubscript{{$\infty$}}} {{Synthesis}}.  {\it IEEE Transactions on Automatic Control. }2006;51(1):71--86.

\bibitem{apkarianMultimodelMultiobjectiveTuning2014}
Apkarian P, Gahinet P, Buhr C. Multi-Model, Multi-Objective Tuning of Fixed-Structure Controllers.  In: 2014 {{European Control Conference}} ({{ECC}}):856--861IEEE; 2014; Strasbourg, France.

\bibitem{RobustControlToolbox}
The~MathWorks {\relax Inc}.. {\it Robust {{Control Toolbox}} Version: 24.2 ({{R2024b}}). } 2025.

\bibitem{doi:10.1049/ip-d.1982.0053}
Doyle J. Analysis of Feedback Systems with Structured Uncertainties.  {\it IEE Proceedings D (Control Theory and Applications). }1982;129(6):242--250.

\bibitem{sanfedinoSatelliteDynamicsToolbox2023}
Sanfedino F, Alazard D, Kassarian E, Somers F. Satellite {{Dynamics Toolbox Library}}: A Tool to Model Multi-Body Space Systems for Robust Control Synthesis and Analysis.  {\it IFAC-PapersOnLine. }2023;56(2):9153--9160.

\bibitem{kwiatkowskiLPVControl2DOF2005}
Kwiatkowski A, Werner H. {{LPV Control}} of a 2-{{DOF Robot Using Parameter Reduction}}.  In: Proceedings of the 44th {{IEEE Conference}} on {{Decision}} and {{Control}}:3369--3374; 2005; Seville, Spain.

\bibitem{rizviKernelBasedPCAApproach2016}
Rizvi SZ, Mohammadpour J, T{\'o}th R, Meskin N. A {{Kernel-Based PCA Approach}} to {{Model Reduction}} of {{Linear Parameter-Varying Systems}}.  {\it IEEE Transactions on Control Systems Technology. }2016;24(5):1883--1891.

\bibitem{koelewijnSchedulingDimensionReduction2020}
Koelewijn PJW, T{\'o}th R. Scheduling {{Dimension Reduction}} of {{LPV Models}} - {{A Deep Neural Network Approach}}.  In: 2020 {{American Control Conference}}:1111--1117; 2020; Denver, US.

\end{thebibliography}
\end{document}